# Integrated nanophotonic platform for on-chip quantum emitter interactions and entanglement

Yinhui Kan[1,2]*, Shailesh Kumar[1], Xujing Liu[1], Antonio I. Fernández-Domínguez[3,4], Sergey I. Bozhevolnyi[1]*

[1]Center for Nano Optics, University of Southern Denmark, DK-5230 Odense M, Denmark.
[2]Niels Bohr Institute, University of Copenhagen, 2100 Copenhagen, Denmark.
[3]Departamento de Física Teórica de la Materia Condensada, Universidad Autónoma de Madrid, E-28049 Madrid, Spain.
[4]Condensed Matter Physics Center (IFIMAC), Universidad Autónoma de Madrid, E-28049 Madrid, Spain.

*Corresponding author. Email: yinhui.kan@nbi.ku.dk; seib@mci.sdu.dk

**Abstract:**

Entanglement between solid-state quantum emitters (QEs) is a key resource for photonic quantum technologies. Achieving such entanglement requires strong and controllable long-range interactions between QEs. However, engineering such coupling remains challenging, particularly for on-chip distant solid-state QEs. Here, we introduce a forward-designed platform that enables ultracompact nanophotonic architectures to mediate enhanced long-range QE–QE interactions via engineered surface plasmon polariton interference. Using this strategy, we realize two distinct configurations: a phase-conjugated elliptic design for energy funneling, and a co-radiating hyperbolic design for its suppression. We experimentally demonstrate large enhancement and suppression of energy transfer rates compared to bare substrates. Furthermore, we predict transient entanglement between spatially separated QEs with concurrence peaking at ~0.493, approaching the theoretical bound in the transient regime. Extending to the multi-QE case, we observe enhanced energy funneling and predict QE-QE entanglement in three-QE configurations. These results establish a compact and scalable framework for on-chip entanglement engineering in integrated quantum nanophotonic systems.

Solid-state quantum emitters (QEs), such as quantum dots, color centres, and molecules, are among the most promising candidates for realizing scalable qubit platforms, owing to their ability to emit single photons upon radiative decay from the excited to the ground state[1–3]. Integrating such emitters into ultracompact, on-chip entangled qubit systems is of significant interest for scalable quantum technologies[4–6]. While short-range entanglement between solid-state emitters can be achieved using high-quality cavities or coherent dipole–dipole interactions[7–11], extending such interactions to longer distances remains a formidable challenge. In particular, mediating the photon-assisted coupling between spatially separated QEs on a chip requires both efficient energy channeling and extremely low radiative damping[12], which is difficult to achieve simultaneously in conventional photonic systems.

Nanophotonic structures, including photonic crystals[13–15], waveguides[16,17], and free-space metasurfaces[18,19], offer a versatile toolbox for mediating photon-assisted interactions between solid-state QEs. However, efficiently and flexibly interfacing these nanostructures with single QEs remains a major bottleneck due to their limited spatial coupling range and fabrication constraints. In contrast to conventional photonic platforms, those realizing QE interactions mediated by surface plasmon polaritons (SPPs) offer a compelling alternative due to a larger freedom in realization of on-chip engineered functionalities[20,21]. Recent works have shown that SPPs launched by QEs can be outcoupled to free-space photons with tailored polarization[22,23] and directionality[24–26]. Yet, it remains elusive how to engineer on-chip environments that would enable efficient emitter-emitter interactions and eventually entanglement mediated by QE-excited propagating SPP waves.

In this work, we propose a forward-designed nanophotonic platform to engineer long-range interactions between on-chip QEs involving interference of QE-launched SPP waves. We design two types of nanostructured configurations: a phase-conjugated elliptic geometry that funnels energy between QEs and a hyperbolic co-radiating geometry that suppresses energy transfer. We experimentally demonstrate significant enhancement and suppression of on-chip energy-transfer rates between nitrogen vacancy centres in nanodiamonds and photon acceptors, validating our design principle. We further model the time evolution of concurrence and predict steady-state and transient entanglement peaking at ~0.493 for emitters separated by several micrometers. Finally, we extend our approach to the multi-QE regime and experimentally validate energy funneling in a three-emitter configuration, paving the way toward scalable on-chip devices enabling SPP-assisted multipartite entanglement.

The integrated nanophotonic platform proposed for realizing on-chip multi-QE interaction (schematically shown in Fig. 1) represents ultracompact forward-designed dielectric surface nanostructures, metasurfaces, integrated with on-chip positioned solid-state QEs. The central idea is to engineer metasurfaces that would enable SPP-mediated energy funneling between spatially separated QEs, thereby realizing efficient and controllable electromagnetic interactions between them. In the configuration considered in this work, each QE presents a vertically (i.e., normal to the chip surface) oriented electric dipole moment (in experiments achieved by a specially configured QE selection[20]) that efficiently and non-radiatively coupled to radially outward-propagating SPPs supported by an underlying planar metallic layer. The use of QE-excited SPPs is favorably compared to that based on conventional photonic modes in unstructured environment: the former opens a way of overcoming both drastic decay of three-dimensionally propagating (far) fields and subwavelength limitations of near-field coupling. By placing QEs at specific locations on the chip surface, the SPPs originating from different QEs can interfere constructively or destructively. We exploit the time-reversed interference pattern to perform the forward metasurface design, thus constructing nanostructures that shape the SPP wavefronts and direct energy funneling from one emitter to others. This approach enables flexible and efficient on-chip energy transfer between distant QEs, thereby significantly enhancing their interaction strength and facilitating multipartite quantum entanglement.

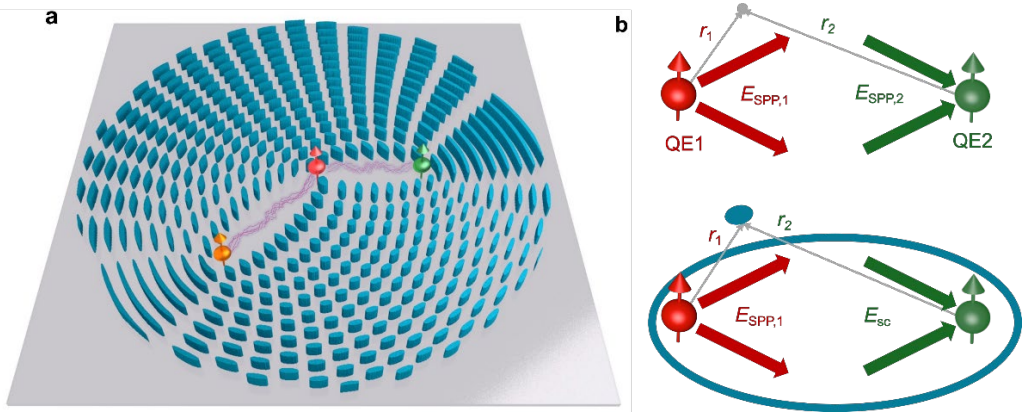

**Figure 1. Integrated nanophotonic platform for on-chip QE interactions and entanglement. a,** Schematic of the on-chip integrated QE- nanophotonic architecture. Vertically oriented QEs radiate via radially outward-propagating SPPs on a metallic surface. The surrounding forward-designed ultracompact metasurface nanostructure reshapes the SPP fields to efficiently channel energy between QEs, enhancing their electromagnetic interactions and enabling multipartite entanglement. **b,** Forward-designed strategy of on-chip holographic metasurface for QE–QE coupling. The recording process (illustrated by the top sub-panel) is based on the interference between predefined reference SPP waves ($E_{SPP,1}$) and signal SPP waves ($E_{SPP,2}$) emitted by two (extendable to multiple) QEs, which can be treated as a source and an acceptor, respectively. The resulting interference pattern (hologram) directly defines the holographic metasurface after the corresponding binarization. In the reconstruction process (illustrated by the bottom sub-panel), when driven by the outgoing SPPs ($E_{SPP,1}$) launched from QE1, the metasurface reconstructs the corresponding incoming SPPs ($E_{sc}$) signal towards QE2.

We first consider the two-QE regime by treating one QE as the radiation source and the other one as the acceptor, both located in the $(x, y)$-plane. The normal to the surface $z$-component of SPP field excited by (and radiating from) the source QE can be described within the $xy$-plane as $E_{SPP,1} \sim \exp(+ik_{SPP}r_{1,xy})$, while that of the converging toward the acceptor as $E_{SPP,2} \sim \exp(-ik_{SPP}r_{2,xy})$, corresponding to a time-reversed configuration. The space-varying part of the resulting SPP intensity interference pattern (Supplementary Note 1) is then given by $I(x,y) \sim \cos[k_{SPP}(r_{1,xy} + r_{2,xy})]$. This term describes concentric ellipses satisfying the condition $r_1 + r_2 = \text{const}$, with focal points at the source and acceptor positions (Fig. 2a). The metasurface nanostructure made of ridges is obtained by binarizing the interference fringes using an appropriate intensity threshold. The threshold is chosen such that the real structure filling ratio $f$ remains close to the preset filling ratio $\xi$ used in generating the interference pattern. To quantify the QE–QE energy transfer[26], we compute the normalized energy transfer rate (*nETR*) defined as $nETR = \frac{\text{Im}[\boldsymbol{\mu}_2^* \cdot \mathbf{G}_c(\mathbf{r}_2,\mathbf{r}_1) \cdot \boldsymbol{\mu}_1]}{\text{Im}[\boldsymbol{\mu}_2^* \cdot \mathbf{G}_s(\mathbf{r}_2,\mathbf{r}_1) \cdot \boldsymbol{\mu}_1]}$, where $\boldsymbol{\mu}_1$ and $\boldsymbol{\mu}_2$ are the QE dipole moment at positions $\mathbf{r}_1$ and $\mathbf{r}_2$, respectively, and $\mathbf{G}_c(\cdot)$, $\mathbf{G}_s(\cdot)$ are the Green's field propagators for the designed metasurface configuration and (bare) substrate, respectively. According to three-dimensional finite-difference time-domain (FDTD) simulations, when the source QE is placed at the left focal point of the elliptic structure, the energy flow is efficiently directed toward the right focal point. The calculated *nETR* distribution in the vicinity of the right focal point demonstrates that our designed metasurface outperforms (in terms of energy transfer rate) the bare substrate by more than three orders of magnitude. To demonstrate the energy transfer suppression (i.e., electromagnetic cloaking of a particular surface point), we design a co-radiating regime, resulting from the pattern corresponding to outward-propagating SPPs from both QEs: $E_{SPP,1} \sim \exp(+ik_{SPP}r_{1,xy})$ and $E_{SPP,2} \sim \exp(+ik_{SPP}r_{2,xy})$. The resulting SPP intensity interference pattern follows then consequently a hyperbolic geometry (Fig. 2b), defined by the condition $r_1 - r_2 = \text{const}$. Following the same binarization and simulation procedure, we compute the *nETR* in the vicinity of the acceptor position and observe a significant suppression of energy transfer, reducing the *nETR* at the acceptor position to ~0.3 relative to the bare substrate.

For the phase-conjugated elliptic design shown in Fig. 1a, varying the distance between QE1 and QE2 ($d_{12}$) leads to the evolution of the generated interference patterns and corresponding metasurfaces, with the QE-connecting line and associated structural constraints emerging and vanishing (Supplementary Note 1, and Fig. S1). One can identify an optimal QE separation that maximizes *nETR* within a given $d_{12}$ range (for example, $d_{12}^{opt} \sim 5.3\lambda_{SPP}$ in the range $5\lambda_{SPP} \leq d_{12} < 6\lambda_{SPP}$), where $\lambda_{SPP}$ denotes the SPP wavelength calculated at the operating wavelength of 670 nm and for a given metasurface filling ratio[25,26]. Furthermore, we propose two approaches for precise control of the resonance wavelength for a given distance between QEs. By varying the width of ellipse ridges (i.e., the metasurface filling ratio and thereby the SPP wavelength), the resonance peak can be tuned continuously in vacuum wavelength (Fig. 2c and Supplementary Fig. S2), which is analogous to controlling the medium refractive index in a Bragg cavity. Second, the resonance wavelength can be engineered through the geometric in-plane scaling of holographic metasurfaces. This approach fundamentally relies on the geometric scaling property of the wave equation in the absence of material dispersion. In our case, as seen in Fig. 2d, the resonance wavelength shifts systematically to shorter (longer) wavelengths under in-plane contraction (expansion). The relative resonance shift $\Delta\lambda/\lambda$ exhibits an approximately linear dependence on the geometric scaling factor (Supplementary Fig. S3), even though the ridge height is kept the same, indicating that its influence is less significant than that of in-plane dimensions. These results demonstrate that our approach enables flexible modulation of energy transfer between on-chip QEs across micrometer-scale distances, thereby outperforming conventional waveguides or V-groove geometries as far as the flexibility and design precision are concerned.

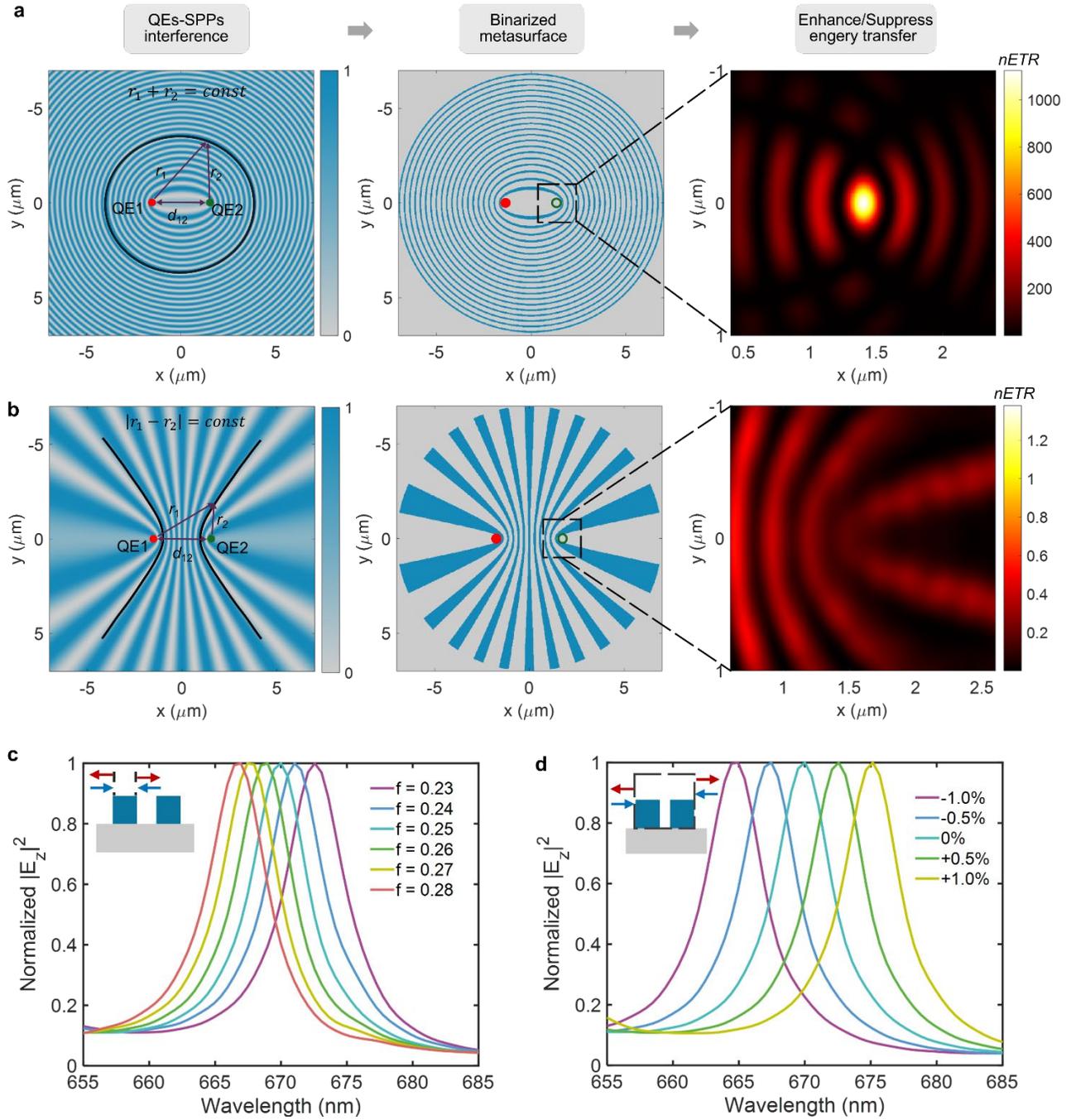

**Figure 2. Forward-designed metasurface nanostructures for on-chip modulating energy transfer between two QEs.** Both vertically oriented QEs are positioned at 30 nm above a substrate, consisting of a 20 nm-thick silica spacer layer atop a silver film, and separated by 2900 nm (about $5.3\lambda_{SPP}$, $\lambda_{SPP}$ is the SPP wavelength) along the *x*-direction. Their vacuum emission wavelengths are set to $\lambda_0 = 670$ nm. The dielectric metasurface nanostructures are designed using a material with the refractive index of 1.41, a height of 175 nm, and the filling ratio of 0.25. **a.** The phase-conjugated elliptic design for enhancement of energy transfer rates. SPP intensity interference pattern is generated by a source (QE, red dot) characterized by an outward-diverging SPP field and acceptor (green dot) characterized by an inward-converging SPP field. The elliptic ridge pattern is obtained via binarization of the interference pattern, with the red dot indicating the QE position. The normalized energy transfer rate (*nETR*) is calculated, showing more than three orders of magnitude enhancement than the bare substrate. **b.** The co-radiating hyperbolic design for suppression of energy transfer rates. SPP intensity interference pattern is formed by outward-propagating SPPs from both QEs. The hyperbolic ridge pattern is obtained via the binarization of the interference pattern. The *nETR* shows ~0.3 relative to the bare substrate. **c.** Precise control of the resonance (field enhancement) wavelength by varying the metasurface filling ratio. **d.** Resonance (field enhancement) wavelength manipulation by geometric in-plane scaling of the holographic metasurface.

To experimentally validate the possibility of modulation of the QE–QE energy transfer via metasurface design, we conceive (as described above) and fabricate metasurface chips containing nanodiamond-embedded nitrogen-vacancy centers (ND-NVs) as solid-state QEs. Three types of configurations are prepared: (*i*) elliptic ridge configuration, (*ii*) control sample without any nanostructure (bare silica-spacer-atop-silver-film substrate), and (*iii*) hyperbolic ridge configuration. The previously developed fabrication process enables precise positioning of the designed metasurface structures, which are electron-beam patterned using hydrogen silsesquioxane (HSQ) as dielectric (negative) photoresist (index $n = 1.41$), with respect to deposited and polarization-emission selected ND-NVs[20,27]. In our experiments, the ND-NVs are positioned at the left focal point of the elliptic or hyperbolic structure, while nanoscale scatterers, acting as acceptors, are fabricated at the corresponding right focal points to scatter incoming SPPs out into detectable far-field photons (Fig. 3a-c).

A tightly focused radially polarized laser beam (532 nm), which provides a strong out-of-plane (*z*) electric-field component, is used to excite ND-NVs, thereby generating a predominantly vertically oriented radiative dipole moment (Supplementary Fig. S4) that primarily couples to radially-propagating SPP fields[20,26]. Fluorescence imaging of the emission patterns (Fig. 3d–f) reveals a bright spot at the right focal point only for the elliptic ridge configuration (Fig. 3d), confirming efficient energy funneling. The out-of-plane scattered emission at the acceptor is dominated by the x polarization (Supplementary Fig. S5), which is parallel to the line connecting the two QEs and to the oscillation direction inside the on-chip cavity. This polarization selection is consistent with the expectation that the field at the acceptor is mainly driven by the SPP fields propagating toward the acceptor directly from the QE as well as those being reflected by the nearest elliptic ridges.

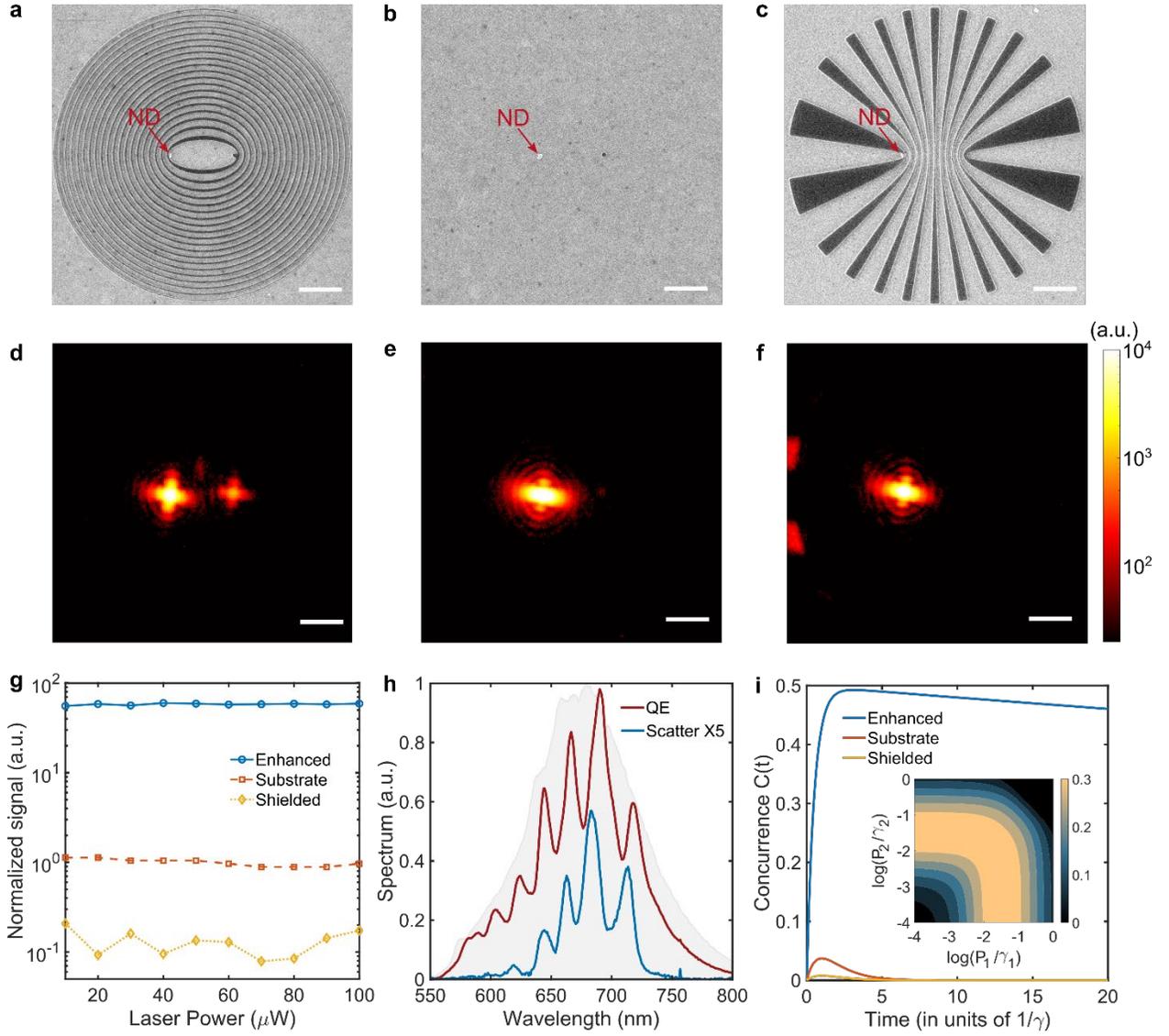

Figure 3. **Experimental demonstration of energy funneling and suppression between solid-state QEs. a-c,** SEM images of fabricated configurations: (a) elliptic ridge configuration, (b) bare substrate, and (c) hyperbolic ridge configuration. ND-NVs are located at the left focal points, and dielectric scatterers are fabricated at the right focal points in all three configurations. **d-f**, Fluorescence images recorded at the image plane under laser ND-NV excitation in the corresponding (a–c) configuration. Scale bars in (a-f) are 2 μm. **g,** Comparison of normalized scattering signals at the right focal points across three configurations. **h,** Emission spectra from the ND-NV and dielectric scatterer in the elliptic configuration (a, d). The gray region denotes a typical ND-NV spectrum measured on a bare substrate. **i,** Simulated time-dependent concurrence for the three corresponding systems. The inset shows steady-state entanglement calculated for the elliptic ridge under continuous incoherent pumping of both QEs (P1, P2 are excitation powers for emitters 1 and 2, respectively.).

In comparison, the signal is significantly reduced on the bare substrate (Fig. 3e) and practically vanishes in the hyperbolic configuration (Fig. 3f). To quantitatively compare the energy transfer efficiency, we measure the scattered emission intensity at the right focal point as a function of excitation power, thus minimizing the influence of background noise. Since each ND-NVs exhibits different intrinsic brightness, the scattered signal is normalized to the emission intensity of the corresponding ND-NVs source. The normalized intensity on the bare substrate is treated as the reference. It is seen that the elliptic configuration exhibits a ~ 60-fold intensity enhancement, while the hyperbolic configuration suppresses the energy transfer down to ~ 0.1 (Fig. 3g). The experimental enhancement factor for the elliptic configuration is found to be lower than the simulated one (Fig. 2a), a difference that can be explained by the influence of various factors. Apart from fabrication-induced deviations from the nominal metasurface parameters, a key factor is that the design was conducted at the specific wavelength of 670 nm, whereas the ND-NV broadband emission was collected in the experimental measurements. Meanwhile, our analysis of the emission spectrum from both the source and the scatterer in the elliptic configuration reveals pronounced periodic oscillations, which are absent in the typical ND-NV spectrum placed on a bare substrate (Fig. 3h). These oscillations indicate the formation (through the metasurface elliptic ridges) of an efficient two-dimensional Bragg cavity for SPP waves travelling back and forth between source and scatterer. The measured free spectral ranges (FSRs) are $22 \pm 2$ nm (source) and $23 \pm 3$ nm (scatterer), values that are consistent with the simulated FSR of 26 nm (Supplementary Fig. S6a). Using the analytic FSR expression $\Delta\lambda = \lambda^2/(2n_g L_{eff})$, where the group index of the SPP mode is estimated as $n_g \approx 1.62$ at 670 nm (for given ridge parameters: height, width and filling ratio), allows one to estimate the effective cavity dimension (along the long ellipse axis): $L_{eff} \cong 5.4$ µm, which corresponds to the metasurface spatial extension (~ 13 elliptic ridges) responsible for 80% of the total *nETR* (Supplementary Fig. S6b). We note that by employing a low-loss substrate, e.g., by reducing the imaginary part of the silver permittivity to Im $\varepsilon \approx 0.1$, one would be able to realize substantially higher peak enhancements and much sharper spectral features (up to a 20-fold increase) compared with the considered realistic lossy case (Supplementary Fig. S7). Ultimately, the maximum achievable *nETR* is also limited because of the unavoidable out-of-plane SPP scattering by the metasurface ridges.

The *nETR* introduced above coincides with the ratio between QE-QE dissipative interaction strengths, $\gamma_{12}$, in the forward-designed and bare-substrate configurations. These emerge in the Born-Markov master equation description of the collective quantum state of the emitters [28], and recently, their maximization has been proposed as a route to achieve high QE entanglement in different nanophotonic devices [16,29]. To explore the realization of on-chip QE interactions and entanglement in our integrated platform, we compute the time-dependent Wootters concurrence [30,31], $C(t)$, which is a well-defined entanglement metric in bipartite systems. Mimicking our experimental setup, we consider a spontaneous emission configuration in which, initially, the source (acceptor) QE is in its excited (ground) state. The time evolution of the concurrence is given by: $C(t) = \sqrt{[\rho_{++}(t) - \rho_{--}(t)]^2 + 4[\text{Im}(\rho_{+-}(t))]^2}$ [19,32], where

$\rho_{++} = \rho_{++}(0)e^{-(\gamma+\gamma_{12})t}$ and $\rho_{--} = \rho_{--}(0)e^{-(\gamma-\gamma_{12})t}$ are the populations of the symmetric $|+\rangle$ and antisymmetric $|-\rangle$ one-excitation Bell states, respectively, and $\gamma$ is the spontaneous decay rate, identical for both QEs owing to the symmetry of the metasurface nanostructures (Fig. 3a) under permutation of the source and acceptor positions. The imaginary part of the coherence in the density matrix evolves as $\text{Im}(\rho_{+-}) = \rho_{+-}(0)e^{-\gamma t}\sin(2g_{12}t)$, where $g_{12} \propto \text{Re}\{\mathbf{G}(\mathbf{r_2},\mathbf{r_1},\lambda_0)\}$ is the QE-QE coherent coupling rate. Conducting the corresponding calculations, one finds that the elliptic configuration exhibits a pronounced peak in the concurrence at $t \sim 2.5/\gamma$, reaching $C_{\text{max}}=0.493$, which approaches the theoretical concurrence maximum (0.5) for the single-excitation subspace (Fig. 3i) and is roughly half of the value for maximally entangled pure states, $C = 1$. In contrast, the flat-substrate and hyperbolic (shielded) configurations yield significantly lower, negligibly small, peak values of $C_{\text{max}} = 0.037$ and $0.008$, respectively (Fig. 3i). This confirms that the forward-designed structure strongly enhances transient entanglement between spatially separated QEs, outperforming previously proposed platforms, such as free-space metasurfaces[19] and plasmonic waveguides ($C_{\text{max}} \approx 0.4$)[16].

Beyond the transient regime, we also investigate steady-state entanglement under continuous incoherent pumping of both QEs. This emulates another experimentally relevant scenario, namely the continuous off-resonant driving of identical QEs[33]. The steady-state density matrix $\rho$ is computed by solving the Born-Markov Master Equation for $\frac{d\rho}{dt} = 0$, with the addition of the pumping Lindbladian term [29]. The steady-state concurrence can be evaluated in terms of the density matrix entries as $C = 2\max\{0, \sqrt{(\rho_{++} - \rho_{--})^2 + 4\text{Im}\{\rho_{+-}\}^2} - \sqrt{\rho_{00}\rho_{33}}\}$, where $\rho_{00}$ and $\rho_{33}$ are the populations of the ground and doubly excited states, respectively. The normalized steady state concurrence as a function of incoherent pumping rates for the two emitters in the elliptic configuration is maximized for certain pumping levels with respect to the dissipative coupling strengths (inset in Fig. 3i). Importantly, the elliptic configuration supports the steady-state concurrence reaching up to $C = 0.3$, while both free-space and flat-substrate platforms yield $C = 0$ (Supplementary Figs. S8-S10). These results highlight the qualitatively distinct performance of our integrated nanophotonic platform with respect to realizing on-chip qubit interactions and entanglement by enabling the formation of substantial entanglement between distant QEs that would remain entirely disentangled in absence of the metasurface.

We further extend our design approach to the multi-QE regime, using a representative configuration of three spatially separated quantum emitters. In this case, the source QE excites a radially propagating SPP field of the form $E_{SPP,1} \sim \exp(+ik_{SPP}r_{1,xy})$ while the two target QEs (acceptors) generate phase-conjugated fields $E_{SPP,2} \sim \exp(-ik_{SPP}r_{2,xy})$ and $E_{SPP,3} \sim \exp(-ik_{SPP}r_{3,xy})$. We design two different three-QE configurations with emitter separations $d_{12} = 4.5\lambda_{SPP}$, $d_{13} = 5.5\lambda_{SPP}$, and $d_{23} = 9\lambda_{SPP}$ (configuration A) and $d_{12} = 3.5\lambda_{SPP}$, $d_{13} = 5.5\lambda_{SPP}$, and $d_{23} = 7\lambda_{SPP}$ (configuration B), as shown in Figs. 4a and 4e. The total interference pattern is constructed as the superposition of two pairwise interference patterns, namely those formed by QE1–QE2 and QE1–QE3, resulting in a multi-focal phase-conjugated regime (Supplementary Note 2; Fig. 4a and 4e). Here, the separations $d_{12}$ and $d_{13}$ are chosen as $(m + 1/2)\lambda_{\text{SPP}}$, while $d_{23}$ is set to $m\lambda_{\text{SPP}}$, with integer $m$. The idea is that pairwise synchronization, QE1–QE2 and QE1–QE3, will be achieved by the corresponding elliptic Bragg cavities as considered above, while SPP fields originated at QE2 (QE3) and propagating toward QE3 (QE2) would still interfere constructively at QE locations with those formed in the corresponding Bragg cavities. We should also mention that the addition of the Q2-Q3 cavity turned out to be very problematic from the viewpoint of maximizing SPP-mediated energy transfer. The reason is that, in the absence of symmetry found in the case with two emitters, the spectral maxima of energy transfer are slightly displaced with respect to each other even without the presence of a third cavity (Supplementary Figs. S11 and S12). The observed differences are fundamentally related to the regime of multiple SPP scattering supported by elliptic Bragg cavities, whereas the holographic metasurface design is based on the single scattering regime (Supplementary Notes 1 and 2). At any rate, the spectral energy transfer maxima can always be fine tuned using the approaches discussed above (Figs. 2c and 2d).

The metasurface designs are obtained by binarizing the total intensity interference patterns with an optimized threshold. We thereby extract the nanostructure layout for each configuration, adding on the corresponding nanoscatterers, and pattern it around pre-located ND-NVs following the same fabrication process as in two-QE designs (Figs. 4b and 4f). The far-field NV emission patterns from the nanoscatterers clearly reveal enhanced energy transfer to both acceptor sites (Figs. 4c and 4g). This indicates the occurrence of multi-QE energy funneling effect. As expected, the *nETR* enhancements in the three-QE configurations are lower than that observed in the two-QE case. This reduction primarily stems from the increased architectural complexity, which makes the individual Bragg cavities leakier as the SPP interaction with metasurface features becomes more susceptible to out-of-plane scattering losses. Nevertheless, our results demonstrate the versatility and flexibility of our design approach for engineering multi-QE interactions, beyond the capabilities of conventional waveguides and cavities. Moreover, the resonance peaks in three-QE design can also be precisely manipulated by geometric in-plane scaling (Supplementary Figs. S11 and S12), which indicates that our multi-QE on-chip platform is compatible with integrated solid-state quantum architectures and can serve as a building block for entanglement generation and state transfer in larger quantum networks[34,35].

To explore theoretically the formation of entanglement among the QEs in the three-QE platforms, we solve (similarly to the bipartite case) the steady-state Born-Markov master equation for the system, parameterized through FDTD simulations in this more complex and computationally demanding geometry. To mimic more closely the experimental condition, we focus now on the incoherent pumping of only the QE1. Thus, using the same theoretical tools as employed in the two-QE case, we compute the steady-state concurrence of the partial density matrices obtained by tracing out one of the qubits in the system. This way, we can reveal the formation of bipartite entanglement. The simulation results obtained for the two three-QE samples (Figs. 4d and h) show that, in all cases, the concurrence peaks up at moderate pumping strengths, yielding up to $C \approx 0.14$ for the qubit pair formed by the central QE (1) and the closest one (2), whose quantum state is given by $\rho_{12}$. The concurrence for the two qubit pairs involving QE3 are smaller but persisting to exist up to certain pumping strength, when the concurrence vanishes. Calculations for the bare substrate reveal that $C = 0$ for all qubit pairs and pumping strengths, which further confirms the remarkable ability of our forward-designed architectures to generate entanglement between distant QEs.

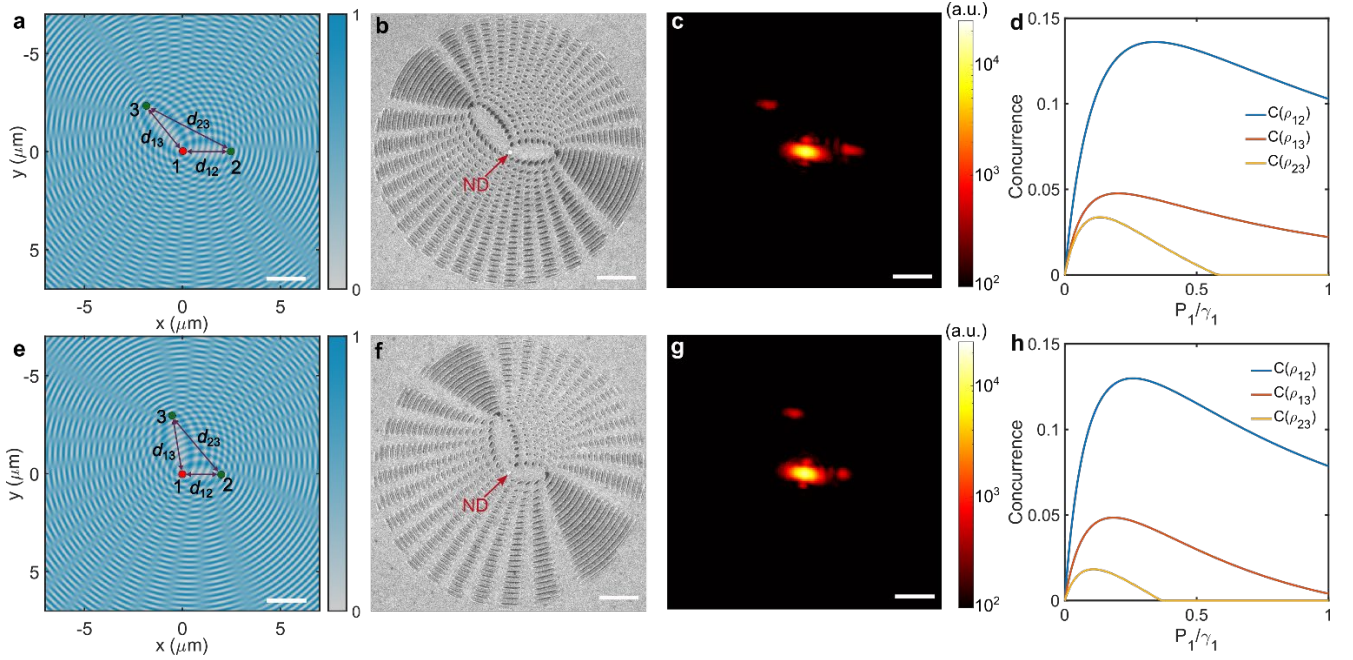

**Figure 4. Multi-QE interference design for enhanced energy funneling. a, e,** Multi-QE phase-conjugated SPP interference patterns: (a) $d_1 = 4.5\lambda_{SPP}$, $d_2 = 5.5\lambda_{SPP}$, $d_3 = 9\lambda_{SPP}$; (e) $d_1 = 3.5\lambda_{SPP}$, $d_2 = 5.5\lambda_{SPP}$, $d_3 = 7\lambda_{SPP}$. **b, f,** SEM images of the fabricated nanostructures patterned around ND-NVs, including nanoscatterers positioned at prospective locations of other QEs to be entangled. **c, g,** Fluorescence emission patterns observed at the image plane for samples (b) and (f), respectively, showing the emission (SPP-mediated) scattering from the nanoscatterers fabricated at the focal points. Scale bars: 2 µm. **d,** Theoretical steady-state concurrence as a function of the pumping strength of the central QE (labelled as 1) for the three qubit-pairs that result from tracing out one of the qubits in the configuration shown in (b). **h,** Same as (d) but for the configuration shown in (f).

In summary, we have introduced a conceptually novel way of generating quantum entanglement between on-chip solid-state QEs and presented the corresponding integrated nanophotonic platform for mediating long-range metasurface-empowered QE interactions. Our approach leverages QE-excited SPPs and engineered interference patterns to create ultracompact metasurface nanostructures that funnel energy between spatially separated QEs with over three orders of magnitude enhancement compared to planar (unstructured) substrates. We experimentally demonstrate this energy transfer enhancement and suppression using ND-NVs integrated with elliptic and hyperbolic metasurfaces, respectively. The engineered QE environment also enables strong quantum correlations, with a maximum transient concurrence evaluated to reach ~0.49, a very high level that approaches the theoretical limit of 0.5 for two qubits in the single-excitation subspace.

Beyond two-QE systems, we extend our forward-design strategy to multi-QE configurations and verify enhanced energy transfer in three-QE geometries. This approach offers clear advantages over cavity- or waveguide-mediated platforms, including high spatial design freedom, on-chip compatibility with nanoscale emitters, and potential for broadband operation unconstrained by narrowband resonances. Moreover, we theoretically verify that our forward-designed metasurfaces yield significant steady-state bipartite entanglement in the three-QE quantum state. The generality and scalability of this approach make it applicable to various types of QEs, such as quantum dots or molecules, and adaptable across different operating wavelengths [1,36,37].

Looking ahead, our platform opens a pathway toward chip-integrated quantum photonic circuits with programmable interaction topologies[38,39]. By incorporating outcoupling metasurfaces in the design process, it may further enable on-demand generation of entangled photon states with designed polarization, spin–orbit, or orbital angular momentum characteristics[40,41]. Recent further advances in metasurface-enabled generation of quantum structured (high-dimensional) light [23], including skyrmions[42], suggest the fascinating possibilities opened with our platform for on-chip generation of nonlocal skyrmions, representing topologically resilient quantum entangled states of light[43]. Our work thus establishes a solid fundament for further exploitations of the presented metasurface-empowered nanophotonic platform, particularly for scalable entanglement engineering in emerging quantum networks and photonic quantum processors[44–46].

## Methods

**Numerical simulations.** Numerical simulations were performed using three-dimensional finite-difference time-domain (FDTD) methods to characterize the interaction between quantum emitters (QEs) and designed holographic nanostructures. The QEs were modeled as vertically oriented electric dipoles operating at a wavelength of 670 nm, corresponding to the fluorescence emission peak region of nitrogen-vacancy centres in nanodiamonds (ND-NVs) at room temperature. The dipole was placed 30 nm above the substrate, which consists of a 20 nm-thick silica spacer atop a 150 nm-thick silver film. The surrounding nanostructures were modeled using a dielectric material with a refractive index of 1.41 and a height of 175 nm. To evaluate the interaction strength, we placed a two-dimensional planar field monitor in the *x–y* plane at the same vertical position as the dipole, located at the target (acceptor) position. The electric fields recorded by this monitor were used to compute the normalized energy transfer rate and extract dyadic Green's functions for both the nanostructured and bare-substrate environments.

**Concurrence calculations.**

*Transient concurrence dynamics.* The transient entanglement dynamics between two quantum emitters are modelled using an unpumped two-qubit master-equation formalism,

$$\dot{\rho} = -\frac{i}{\hbar}\left[-\hbar g_{12}(\sigma_1^\dagger \sigma_2 + \sigma_2^\dagger \sigma_1), \rho\right] + \sum_{\alpha,\beta=1,2} \frac{\gamma_{\alpha\beta}}{2}(2\sigma_\beta \rho \sigma_\alpha^\dagger - \sigma_\alpha^\dagger \sigma_\beta \rho - \rho \sigma_\alpha^\dagger \sigma_\beta),$$

The collective coupling parameters are extracted from the FDTD-simulated electric field as

$$\gamma_{\alpha\beta} \propto \mathrm{Im}\left(G_{zz}(r_\alpha, r_\beta)\right), g_{\alpha\beta} \propto \frac{1}{2}\mathrm{Re}\left(G_{zz}(r_\alpha, r_\beta)\right)$$

where $G_{zz}$ denotes the *zz*-entry of the dyadic Green's function at the QE locations.

For the initial condition $|e_1 g_2\rangle$, the density matrix entries within the single-excitation manifold read

$$\rho_{++}(t) = \frac{1}{2}e^{-(\gamma+\gamma_{12})t}, \quad \rho_{--}(t) = \frac{1}{2}e^{-(\gamma-\gamma_{12})t}, \quad \mathrm{Im}\rho_{\pm}(t) = \frac{1}{2}e^{-\gamma t}\sin(2g_{12}t)$$

expressed in the Bell state basis $|\pm\rangle = \frac{1}{\sqrt{2}}[|e_1 g_2\rangle \pm |g_1 e_2\rangle]$.

The time-dependent concurrence is given by

$$C(t) = \sqrt{[\rho_{++}(t) - \rho_{--}(t)]^2 + 4[\mathrm{Im}\rho_{\pm}(t)]^2}.$$

*Steady-state concurrence.* The steady-state entanglement in two and three QEs under incoherent pumping can be described using a generalization of the master equation employed in the transient analysis of two QEs, by including the pumping terms $P_\alpha$, we can write

$$\dot{\rho} = -\frac{i}{\hbar}[\widehat{H},\rho] + \sum_{\alpha,\beta}^{N} \frac{\gamma_{\alpha\beta}}{2} L_{\hat{\sigma}_\alpha^\dagger,\hat{\sigma}_\beta}\{\rho\} + \sum_{\alpha}^{N} \frac{P_\alpha}{2} L_{\hat{\sigma}_\alpha,\hat{\sigma}_\alpha^\dagger}\{\rho\} = 0,$$

where the Lindblad super-operators are defined as

$$\mathcal{L}_{\hat{O}\hat{O}'}\{\rho\} = 2\hat{O}'\rho\,\hat{O} - \{\hat{O}\hat{O}',\rho\}$$

and the coherent and dissipative couplings have the same form as above. Solving the stationary condition for two identical emitters, the steady-state density matrix

$$\rho = \begin{pmatrix} \rho_{00} & 0 & 0 & 0 \\ 0 & \rho_{11} & \rho_{12} & 0 \\ 0 & \rho_{12}^* & \rho_{22} & 0 \\ 0 & 0 & 0 & \rho_{33} \end{pmatrix}$$

is obtained. From this density matrix, the concurrence is calculated as

$$C(\rho) = 2\max\{0, |\rho_{12}| - \sqrt{\rho_{00}\rho_{33}}\} = 2\max\{0, \sqrt{(\rho_{++} - \rho_{--})^2 + 4\mathrm{Im}\{\rho_{+-}\}^2} - \sqrt{\rho_{00}\rho_{33}}\}.$$

where, in the right-hand side of the equation, the Bell state basis has been employed in the first-excitation manifold.

The steady-state density matrix for three QEs has a (1-3-3-1) block diagonal form similar to the two-qubit one above. In fact, after the tracing of one of the emitters, here labelled as $\gamma$, the partial density matrices for the remaining QEs can be obtained, $\rho_{\alpha\beta} = \mathrm{tr}_\gamma\{\rho\}$. These have exactly the same form as the two-QE matrix above.

**Device fabrication.** The QE-integrated nanophotonic structures were fabricated using a two-step aligned electron-beam lithography (EBL) process. First, a 150 nm-thick silver film was thermally evaporated onto a silicon wafer, followed by deposition of a 20 nm $SiO_2$ layer via magnetron sputtering, forming a low-loss plasmonic substrate. Gold alignment markers (35 nm thick) were defined through standard EBL, metal evaporation, and lift-off. ND-NVs were then spin-coated onto the substrate at 3000 rpm for 60 s from a diluted dispersion. Their positions were identified using dark-field optical microscopy, enabling precise emitter registration relative to the gold markers. A hydrogen silsesquioxane (HSQ) resist layer (~175 nm) was spin-coated at 3000 rpm for 45 s and soft-baked at 160 °C for 2 min. The designed holographic nanostructures were patterned via a second, marker-aligned EBL step. The resist was developed in a 25% tetramethylammonium hydroxide (TMAH) solution for 4 min, followed by rinsing in deionized water for 60 s.

**Optical characterization.** The QE-integrated nanostructures were optically characterized using a confocal microscopy setup schematically shown in Supplementary Fig. S4. A radially polarized continuous-wave laser at 532 nm was used to excite ND-NVs, preferentially driving dipole transitions oriented normal to the substrate. Both excitation and collection were performed through a high-numerical-aperture objective (Olympus MPLFLN 100×, NA = 0.9). Fluorescence imaging was carried out by scanning the sample with a piezoelectric stage while the collected emission was directed to a camara and avalanche photodiodes (APDs). A galvanometric mirror and confocal pinhole were used to spatially filter the emission, allowing selective collection of light from individual quantum emitters and nanostructure-mediated scatterers. Spectral measurements were performed using a fibre-coupled spectrometer (QE Pro, Ocean Optics) in the 550–800 nm wavelength range.

**Acknowledgements**

The authors acknowledge the support from European Union's Horizon Europe research and innovation programme under the Marie Skłodowska-Curie Action (Grant agreement No. 101064471) (Y.H.K.), Villum Kann Rasmussen Foundation (Award in Technical and Natural Sciences 2019) (S.I.B.), Otto Mønsteds Fond and the Spanish Ministry of Science, Innovation and Universities (Grant No. PID2024-156077OB-I00) (A.I.F.D.).

**Author contributions**

S.I.B. and Y.H.K. conceived the idea. Y.H.K., A.I.F.D., and S.I.B. performed theoretical modelling. Y.H.K. and X.J.L. fabricated samples. Y.H.K., S.K., and X.J.L. performed experimental measurement. Y.H.K. and A.I.F.D. performed calculation of concurrence. Y.H.K., S.K., X.J.L., and S.I.B. analysed the data. S.I.B. and Y.H.K. supervised the project. Y.H.K. and S.I.B. wrote the manuscript with contributions from all authors.

**Competing interests**

The authors declare no competing interests.

# Supporting Information

# Integrated nanophotonic platform for on-chip quantum emitter interactions and entanglement

Yinhui Kan[1,2]*, Shailesh Kumar[1], Xujing Liu[1], Antonio I. Fernández-Domínguez[3,4], Sergey I. Bozhevolnyi[1]*

[1]Center for Nano Optics, University of Southern Denmark, DK-5230 Odense M, Denmark.
[2]Niels Bohr Institute, University of Copenhagen, 2100 Copenhagen, Denmark.
[3]Departamento de Física Teórica de la Materia Condensada, Universidad Autónoma de Madrid, E-28049 Madrid, Spain.
[4]Condensed Matter Physics Center (IFIMAC), Universidad Autónoma de Madrid, E-28049 Madrid, Spain.
*Corresponding author. Email: yinhui.kan@nbi.ku.dk; seib@mci.sdu.dk

## Note 1 Design principle of on-chip holographic metasurface for QE–QE coupling

The on-chip holographic metasurface is designed based on the interference between two surface plasmon polariton (SPP) waves associated with two spatially separated quantum emitters (QEs). The goal is to engineer a metasurface that converts the SPP field emitted by one QE into a converging SPP field focused onto the position of the second QE, thereby enabling controlled QE–QE coupling mediated by surface plasmons.

### Interference of reference and signal SPP waves

We consider two QEs located at positions labelled as point 1 and point 2. Reference SPP waves are generated by QE 1 and propagates radially along the metal surface. The fields at an arbitrary in-plane position $(x, y)$ can be written as

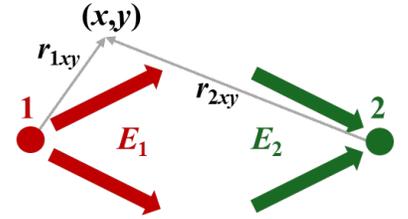

$$E_1 = E_1^0 \exp(+ik_{\text{spp}} r_{1xy}) \tag{S1}$$

where $r_{1xy}$ denotes the in-plane distance between point 1 and $(x, y)$, and $k_{\text{spp}}$ is the SPP wavevector. The signal waves are defined as converging SPP waves directed toward point 2, described by

$$E_2 = E_2^0 \exp(-ik_{\text{spp}} r_{2xy}) \tag{S2}$$

with $r_{2xy}$ being the distance from point 2 to $(x, y)$. The opposite signs in the phase factors reflect the outgoing and incoming nature of the two SPP waves, respectively.

The interference intensity distribution formed by these two SPP waves is given by

$$I = |\boldsymbol{E}_1 + \boldsymbol{E}_2|^2 = \left| \boldsymbol{E}_1^0 \exp(+ik_{spp} r_{1xy}) + \boldsymbol{E}_2^0 \exp(-ik_{spp} r_{2xy}) \right|^2$$
$$= I_1^0 + I_2^0 + \boldsymbol{E}_1^0 \cdot \boldsymbol{E}_2^{0*} \exp[+i(k_{spp} r_{1xy} + k_{spp} r_{2xy})] + \boldsymbol{E}_1^{0*} \cdot \boldsymbol{E}_2^0 \exp[-i(k_{spp} r_{2xy} + k_{spp} r_{1xy})] \tag{S3}$$

The interference term depends on the sum $r_{1xy} + r_{2xy}$, indicating that the contours of constant phase satisfy $k_{\text{spp}} r_{1xy} + k_{\text{spp}} r_{2xy} = \text{const}$, or $r_{1xy} + r_{2xy} = \text{const}$. As a result, metasurface ridges defined by the interference fringes take the form of concentric ellipses, with the two points acting as the foci.

### Reconstruction of the converging SPP waves

When the holographic metasurface is illuminated by the reference SPP waves emitted from point 1, the scattered field can be expressed as

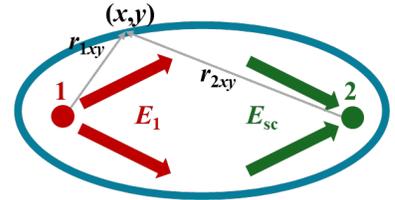

$$\boldsymbol{E}_{sc}(x_i, y_i) \sim I \boldsymbol{E}_1$$
$$= (I_1^0 + I_2^0)\boldsymbol{E}_1^0 \exp(+ik_{spp} r_{1xy}) + (\boldsymbol{E}_1^0 \cdot \boldsymbol{E}_2^{0*})\boldsymbol{E}_1^0 \exp[-i(2k_{spp} r_{1xy} + k_{spp} r_{2xy})]$$
$$+ (\boldsymbol{E}_1^{0*} \cdot \boldsymbol{E}_2^0)\boldsymbol{E}_1^0 \exp[-ik_{spp} r_{2xy}] \tag{S4}$$

The first term in the above expression represents the QE-excited outgoing (from point 1) SPP waves; The second term generates rapidly oscillating electromagnetic fields that are not supported by the SPP dispersion (apart from the line connecting the two points), and the last term reconstructs the converging (to point 2) signal SPP waves.

When the structure is illuminated by the outgoing SPP waves emitted from point 2, the scattered field takes an analogous form

$$E_{sc}(x_i, y_i) \sim IE_{2div}$$
$$= (I_1^0 + I_2^0)\boldsymbol{E}_2^0 \exp(+ik_{spp}r_{2xy}) + (\boldsymbol{E}_1^0 \cdot \boldsymbol{E}_2^{0*})\boldsymbol{E}_2^0 \exp[+i(2k_{spp}r_{2xy} + k_{spp}r_{1xy})]$$
$$+ (\boldsymbol{E}_1^{0*} \cdot \boldsymbol{E}_2^0)\boldsymbol{E}_2^0 \exp[-ik_{spp}r_{1xy}] \qquad (S5)$$

The first term in the above expression represents the QE-excited outgoing (from point 2) SPP waves, the second one generates too rapid field oscillations to comply with the SPP dispersion, and the last term reconstructs the converging (to point 1) signal SPP waves. This reciprocity arises from the fact that the hologram does not change when both interfering SPP waves are changed to the conjugated ones: $|\boldsymbol{E}|^2 \equiv |\boldsymbol{E}^*|^2$.

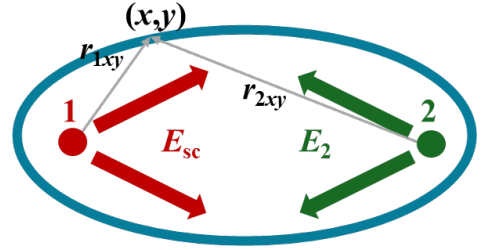

**Interference along the QE-connecting line and structural constraint**

In addition to the general elliptic interference pattern, it is instructive to examine the interference intensity along the straight line connecting the two QEs. This analysis provides a practical criterion for determining whether the QE–QE connecting line should be covered by a metasurface ridge or left as a clean metal surface. We consider the interference intensity along the line $y = 0$, connecting the two QEs and assume normalized field amplitudes $E_1^0 = E_2^0 = 1$. The interference intensity reduces to

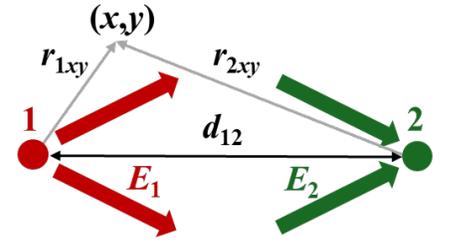

$$I(x, y = 0) = |\boldsymbol{E}_1 + \boldsymbol{E}_2|^2 = 2 + 2\cos(k_{spp}d_{12}) = 2\left[\cos\left(\frac{k_{spp}d_{12}}{2}\right)\right]^2 = 2\left[\cos\left(\frac{\pi d_{12}}{\lambda_{spp}}\right)\right]^2 \qquad (S6)$$

where $d_{12}$ denotes the distance between the two QEs and $\lambda_{spp}$ is the SPP wavelength. This expression reveals that the interference intensity along the QE-connecting line oscillates periodically as a function of the emitter separation. For separations satisfying $d_{12} \approx m\lambda_{spp}$, with integer $m$, constructive interference occurs along the connecting line, corresponding to an intensity maximum. In this case, a holographic ridge would appear along the connecting line as an extreme case of metasurface ellipses.

In contrast, for separations satisfying $d_{12} \approx (m + \frac{1}{2})\lambda_{spp}$, destructive interference occurs along the connecting line, resulting in an intensity minimum. Under this condition, the connecting line naturally remains free of metasurface features, leaving the bare substrate between the two QEs.

This interference-based criterion (illustrated by plots in Fig. S1) provides a simple and physically transparent rule for choosing the appropriate QE separation depending on whether the region directly connecting the two QEs should be patterned or left unstructured, while remaining fully consistent with the global holographic design principle described above. Note that the QE connecting ridge might jeopardize the realization of constructive interference of all SPP waves converging to one or another point due to the differences in the effective propagation constants of the converging waves, because most waves would cross only a few ridges propagating mainly along free (unstructured) metal surface.

**Note 2 Design principle of on-chip holographic metasurface for three-QEs coupling**

In this section, we extend the on-chip holographic metasurface concept to the case of three spatially separated QEs. The goal is to demonstrate that the interference-based holographic design principle naturally generalizes to multi-emitter systems and enables simultaneous, deterministic coupling between multiple remotely located QEs.

We consider three QEs located at positions labeled as points 1, 2, and 3. QE 1 serves as the reference source and excites outgoing SPP waves propagating along the surface. QEs 2 and 3 are treated as target emitters associated with converging signal SPP waves (note that only the out-of-plane electric field component ($E_z$) is considered). The total interference intensity used to define the on-chip holographic metasurface is constructed as the sum of two pairwise interference patterns, 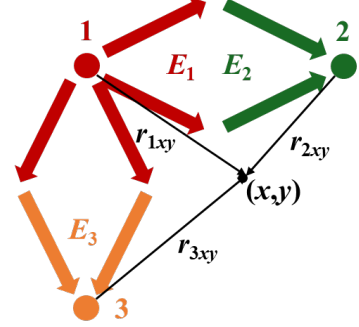

$$I = |E_1 + E_2|^2 + |E_1 + E_3|^2 \tag{S7}$$

which can be expanded as

$$I = I_{tot}^0 + E_1^0 \cdot E_2^{0*} \exp[i(k_{spp}r_{1xy} + k_{spp}r_{2xy})] + E_1^{0*} E_2^0 \exp[-i(k_{spp}r_{2xy} + k_{spp}r_{1xy})]$$
$$+ E_1^0 \cdot E_3^{0*} \exp[i(k_{spp}r_{1xy} + k_{spp}r_{3xy})] + E_1^{0*} \cdot E_3^0 \exp[-i(k_{spp}r_{3xy} + k_{spp}r_{1xy})] \tag{S8}$$

When the holographic metasurface is illuminated by the outgoing SPP waves emitted from QE 1, the scattered field can be written as

$$E_{sc}(x_i, y_i) \sim I E_1$$
$$= I_{tot}^0 E_1^0 \exp(ik_{spp}r_{1xy}) + (E_1^0 \cdot E_2^{0*}) E_1^0 \exp[i(2k_{spp}r_{1xy} + k_{spp}r_{2xy})]$$
$$+ (E_1^{0*} \cdot E_2^0) E_1^0 \exp[-ik_{spp}r_{2xy}] + (E_1^0 \cdot E_3^{0*}) E_1^0 \exp[i(2k_{spp}r_{1xy} + k_{spp}r_{3xy})]$$
$$+ (E_1^{0*} \cdot E_3^0) E_1^0 \exp[-ik_{spp}r_{3xy}] \tag{S9}$$

The first term represents the QE-excited outgoing (from point 1) SPP reference waves; the second and fourth terms generate rapidly oscillating fields that are not supported by the configuration(as explained above); the third and fifth terms reconstruct the converging (to points 2 and 3) signal SPP waves, respectively; the fourth term generates the outgoing from (point 3) SPP waves. The holographic metasurface constructed in this manner is invariant under asynchronous exchange of convergence and divergence within each reference–signal pair. Consequently, flipping the roles of ingoing and outgoing SPP waves within two pairs does not modify the recorded hologram.

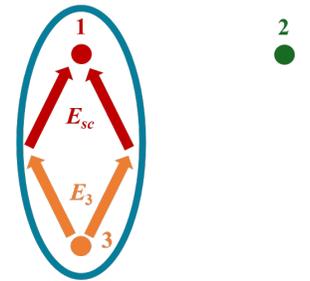

In addition to illumination by the reference emitter QE 1, the response of the holographic metasurface can be analyzed when it is excited by one of the target emitters. For instance, when considering illumination by the SPP waves emitted from QE 3, the scattered field can be written as

$$\begin{aligned}
\boldsymbol{E}_{sc}(x_i, y_i) &\sim I\boldsymbol{E}_1 \\
&= I_{tot}^0 \boldsymbol{E}_3^0 \exp(ik_{spp}r_{3xy}) + (\boldsymbol{E}_1^0 \cdot \boldsymbol{E}_2^{0*})\boldsymbol{E}_3^0 \exp[ik_{spp}(r_{1xy} + r_{2xy} + r_{3xy})] \\
&+ (\boldsymbol{E}_1^{0*} \cdot \boldsymbol{E}_2^0)\boldsymbol{E}_3^0 \exp[-ik_{spp}(r_{1xy} + r_{2xy} - r_{3xy})] + (\boldsymbol{E}_1^0 \\
&\cdot \boldsymbol{E}_3^{0*})\boldsymbol{E}_3^0 \exp[i(2k_{spp}r_{3xy} + k_{spp}r_{1xy})] + (\boldsymbol{E}_1^{0*} \cdot \boldsymbol{E}_3^0)\, \boldsymbol{E}_3^0 \exp[-ik_{spp}r_{1xy}]
\end{aligned} \quad (S10)$$

The first term in the above expression represents the QE-excited outgoing (from point 3) SPP waves; the second, third and fourth terms represent rapidly oscillating fields not supported by the configuration; the fifth term reconstructs the converging (to point 1) SPP waves. A similar situation takes place when the metasurface is illuminated with the QE-excited radiated from point 2 SPP waves.

**Note 3 Calculation of the Time-Dependent Concurrence**

To evaluate the transient entanglement dynamics between two spatially separated QEs, we calculate the concurrence $C(t)$ from the time evolution of their collective Bell states. The analysis follows the standard two-qubit master-equation framework [1, 2].

For two identical two-level emitters with lowering operators $\sigma_{1,2}$ and individual decay rate $\gamma$, the system density matrix $\rho$ obeys the Lindblad master equation:

$$\dot{\rho} = -\frac{i}{\hbar}\left[-\hbar g_{12}(\sigma_1^\dagger \sigma_2 + \sigma_2^\dagger \sigma_1), \rho\right] + \sum_{\alpha,\beta=1,2} \frac{\gamma_{\alpha\beta}}{2}\left(2\sigma_\beta \rho \sigma_\alpha^\dagger - \sigma_\alpha^\dagger \sigma_\beta \rho - \rho \sigma_\alpha^\dagger \sigma_\beta\right), \quad (S11)$$

where $\gamma_{1,2}$ and $g_{12}$ denote, respectively, the collective dissipative and coherent coupling rates between the emitters, which are defined as the imaginary and real parts of the dyadic Green's tensor:

$$g_{12} \propto \mathrm{Re}G(\boldsymbol{r}_i, \boldsymbol{r}_j, \omega_0), \qquad \gamma_{12} \propto \mathrm{Im}G(\boldsymbol{r}_i, \boldsymbol{r}_j, \omega_0), \quad (S12)$$

In the single-excitation subspace, defined by the symmetric and antisymmetric Bell states $|+\rangle = (|e_1 g_2\rangle + |g_1 e_2\rangle)/\sqrt{2}$ and $|-\rangle = (|e_1 g_2\rangle - |g_1 e_2\rangle)/\sqrt{2}$, the time-dependent concurrence is:

$$C(t) = \sqrt{[\rho_{++}(t) - \rho_{--}(t)]^2 + 4[\mathrm{Im}\rho_\pm(t)]^2}, \quad (S13)$$

with

$$\rho_{++}(t) = \frac{1}{2}e^{-(\gamma+\gamma_{12})t}, \qquad \rho_{--}(t) = \frac{1}{2}e^{-(\gamma-\gamma_{12})t}, \qquad \mathrm{Im}\rho_\pm(t) = \frac{1}{2}e^{-\gamma t}\sin(2g_{12}t), \quad (S14)$$

Equations (S3)–(S4) correspond to the situation where the source emitter is initially excited while the acceptor remains in the ground state. The collective parameters $\gamma_{12}$ and $g_{12}$ are obtained from the simulated local electric field of one emitter (source) acting on the other (acceptor). For dipoles polarized along z:

$$\zeta = \frac{E_z^{(2)}}{\mathrm{Im}E_z^{(1)}}, \qquad \gamma_{12} = \mathrm{Im}(\zeta), \qquad g_{12} = -\frac{1}{2}\mathrm{Re}(\zeta), \quad (S15)$$

where $E_z^{(1)}$ and $E_z^{(2)}$ are the z-components of the electric field at the source and acceptor positions, respectively. In what follows, time is expressed in units of $1/\gamma$ so that the single-emitter decay rate serves as the reference.

The time-dependent concurrence is then evaluated as:

$$C(t) = \sqrt{\left[\frac{1}{2}e^{-(1+\tilde{\gamma}_{12})t} - \frac{1}{2}e^{-(1-\tilde{\gamma}_{12})t}\right]^2 + [e^{-t}\sin(2\tilde{g}_{12}t)]^2}, \tag{S16}$$

where $\tilde{\gamma}_{12} = \gamma_{12}/\gamma$ and $\tilde{g}_{12} = g_{12}/\gamma$. In practice, the complex ratio $\frac{E_z^{(2)}}{\mathrm{Im}E_z^{(1)}}$ is extracted directly from the simulations for each configuration - elliptic (enhanced), flat-substrate, and (hyperbolic) shielded. The resulting $C(t)$ curves reproduce the distinct temporal behaviors plotted in Fig. 3i of the main text. The imaginary part of the normalized field ratio governs the collective decay $\gamma_{12}$, responsible for dissipative entanglement, while the real part yields the coherent exchange rate $g_{12}$. A large $\gamma_{12}$ and small $g_{12}$ combination maximizes $C_{\max}$, with the theoretical limit $C_{\max} = 0.5$ reached for $\gamma_{12} = \gamma$ and $g_{12} = 0$. Our inverse-designed structure achieves $C_{\max} = 0.493$, approaching this fundamental bound.

**Note 4 Steady-State Density-Matrix Analysis and Concurrence Calculation**

To complement the transient dynamics, the steady-state entanglement under incoherent pumping were also investigated. Following the work by Düng, Knöll and Welsch [3], the master equation describing the photon-induced interactions between an ensemble of $N$ QEs in the presence of the same photon vacuum has the form

$$\dot{\rho} = -\frac{i}{\hbar}[\hat{H}, \rho] + \sum_{i,j}^{N} \frac{\gamma_{ij}}{2} L_{\hat{\sigma}_i^\dagger, \hat{\sigma}_j}\{\rho\} + \sum_{i}^{N} \frac{P_i}{2} L_{\hat{\sigma}_i, \hat{\sigma}_i^\dagger}\{\rho\}, \tag{S17}$$

where the Hamiltonian,

$$\hat{H} = \sum_{i=1}^{N} \hbar\omega_i \hat{\sigma}_i^\dagger \hat{\sigma}_i + \sum_{i \neq j=1}^{N} \hbar g_{ij}\, \hat{\sigma}_i^\dagger \hat{\sigma}_j, \tag{S18}$$

describes the coherent interaction between the QEs and the unitary evolution of the system, and the Lindblad super-operators,

$$\mathcal{L}_{\hat{O}\hat{O}'}\{\rho\} = 2\hat{O}'\rho\,\hat{O} - \{\hat{O}\hat{O}', \rho\}, \tag{S19}$$

account for the spontaneous decay and incoherent interactions between the QEs, as well as their incoherent pumping, for example, by a strongly detuned laser. The creation and annhiliation operators for the QE excitons satisfy $[\hat{\sigma}_i, \hat{\sigma}_j^\dagger] = \delta_{ij}(1 - 2\hat{\sigma}_i^\dagger \hat{\sigma}_i)$ and $[\hat{\sigma}_i, \hat{\sigma}_j] = 0$.

Conveniently, using the Born-Markov approximation, the parameters in Eq. (S17) can be expressed in terms of the electromagnetic Dyadic Green's function as

$$g_{ij} = \frac{\omega_0^2}{\pi\varepsilon_0\hbar c^2} \boldsymbol{\mu}_i \, \text{Re}\{\mathbf{G}(\mathbf{r}_i, \mathbf{r}_j, \omega_0)\} \boldsymbol{\mu}_j, \tag{S20}$$

$$\gamma_{ij} = \frac{2\omega_0^2}{\pi\varepsilon_0\hbar c^2} \boldsymbol{\mu}_i \, \text{Im}\{\mathbf{G}(\mathbf{r}_i, \mathbf{r}_j, \omega_0)\} \boldsymbol{\mu}_j, \tag{S21}$$

where $\boldsymbol{\mu}_i$ is the real-valued electric dipole moment of the $i$-th QE. We assume that the QE frequencies are all within a narrow frequency window around a central frequency ($\omega = \omega_0 + \delta_i$ with small $\delta_i$), introduced to evaluate the Green's functions above. Using a three-level-system description of the QEs, with an upper level greatly detuned from the excitons described by the $\hat{\sigma}$ operators above, the incoherent pumping rate can be expressed as [4]

$$P_i = \frac{4\tilde{\Omega}_i^2}{\tilde{\Gamma}_i}, \tag{S22}$$

where $\tilde{\Omega}_i = \mathbf{E}_{\text{pump}}(\mathbf{r}_i, \omega_{\text{pump}}) \cdot \tilde{\boldsymbol{\mu}}_i$ is the product of the pump field times the dipole moment of the transition between the ground and the detuned level, and $\tilde{\Gamma}_i$ is the fast decay rate between this state and the excited state of the remaining two-level system. This expression for the incoherent pumping rate is obtained under the adiabatic elimination of the third state assuming that $1/\tilde{\Gamma}_i$ is the shortest time in the internal dynamics of the three-level system.

To solve Eq. (S17), and resolve the quantum dynamics of the density matrix of the QE ensemble, we simply project the master equation over the different elements of the Hilbert basis, having

$$\dot{\rho}_{\alpha\beta} = \left\langle \alpha \left| \frac{d}{dt}\rho \right| \beta \right\rangle = -\frac{i}{\hbar} \langle \alpha|[\hat{H},\rho]|\beta\rangle + \sum_{i,j}^{N} \frac{\gamma_{ij}}{2} \left\langle \alpha \left| L_{\hat{\sigma}_i^\dagger, \hat{\sigma}_j}\{\rho\} \right| \beta \right\rangle + \sum_{i}^{N} \frac{P_i}{2} \left\langle \alpha \left| L_{\hat{\sigma}_i, \hat{\sigma}_i^\dagger}\{\rho\} \right| \beta \right\rangle. \tag{S23}$$

If we are interested in the steady state for the system, we get a set of homogeneous equations of the form

$$0 = -\frac{i}{\hbar} \langle \alpha|[\hat{H},\rho^{ss}]|\beta\rangle + \sum_{i,j}^{N} \frac{\gamma_{ij}}{2} \left\langle \alpha \left| L_{\hat{\sigma}_i^\dagger, \hat{\sigma}_j}\{\rho^{ss}\} \right| \beta \right\rangle + \sum_{i}^{N} \frac{P_i}{2} \left\langle \alpha \left| L_{\hat{\sigma}_i, \hat{\sigma}_i^\dagger}\{\rho^{ss}\} \right| \beta \right\rangle, \tag{S24}$$

which need to be complemented by the condition

$$\text{tr}\{\rho^{ss}\} = \sum_\alpha \rho^{ss}_{\alpha\alpha} = 1, \tag{S25}$$

which ensures that the total probability of the QE ensembles occupying any state in the Hilbert space is equal to 1.

For the case of 2 identical QEs ($\omega_1 = \omega_2 = \omega_0$), and labeling the states forming the basis of the Hilbert space as,

$$|0\rangle = |g\rangle_1|g\rangle_2 = |gg\rangle,$$
$$|1\rangle = |e\rangle_1|g\rangle_2 = |eg\rangle,$$
$$|2\rangle = |g\rangle_1|e\rangle_2 = |ge\rangle,$$
$$|3\rangle = |e\rangle_1|e\rangle_2 = |ee\rangle.$$

Eqs. (S24) and (S25) yield a system of 7 linear equations (only 6 of them are independent) of the form

$$\dot{\rho}_{00} = 0 = -(P_1 + P_2)\rho_{00} + \gamma_{12}(\rho_{12} + \rho_{12}^*) + \gamma_{11}\rho_{11} + \gamma_{22}\rho_{22},$$
$$\dot{\rho}_{11} = 0 = 2P_1\rho_{00} + 2ig_{12}(\rho_{12} - \rho_{12}^*) - \gamma_{12}(\rho_{12} + \rho_{12}^*) - 2(\gamma_{11} + P_2)\rho_{11} + 2\gamma_{22}\rho_{33},$$
$$\dot{\rho}_{22} = 0 = 2P_2\rho_{00} - 2ig_{12}(\rho_{12} - \rho_{12}^*) - \gamma_{12}(\rho_{12} + \rho_{12}^*) - 2(\gamma_{22} + P_1)\rho_{22} + 2\gamma_{11}\rho_{33},$$
$$\dot{\rho}_{12} = 0 = 2ig_{12}(\rho_{22} - \rho_{11}) + (\gamma_{11} + P_1 + \gamma_{22} + P_2)\rho_{12} + \gamma_{12}(\rho_{11} + \rho_{22} - 2\rho_{33}), \quad \text{(S26)}$$
$$\dot{\rho}_{12}^* = 0 = 2 - ig_{12}(\rho_{22} - \rho_{11}) + (\gamma_{11} + P_1 + \gamma_{22} + P_2)\rho_{12}^* + \gamma_{12}(\rho_{11} + \rho_{22} - 2\rho_{33})$$
$$\dot{\rho}_{33} = 0 = P_2\rho_{11} + P_1\rho_{22} - (\gamma_{11} + \gamma_{22})\rho_{33},$$
$$\text{tr}\{\rho\} = 1 = \rho_{00} + \rho_{11} + \rho_{22} + \rho_{33}.$$

where we have used that $g_{21} = g_{12}$, $\gamma_{21} = \gamma_{12}$, and that the off-diagonal coherences satisfy $\rho_{21} = \rho_{12}^*$. We can immediately conclude that the density matrix for the quantum state of the QE pair has the form (note that we have dropped the SS superindex for simplicity)

$$\rho = \begin{pmatrix} \rho_{00} & 0 & 0 & 0 \\ 0 & \rho_{11} & \rho_{12} & 0 \\ 0 & \rho_{12}^* & \rho_{22} & 0 \\ 0 & 0 & 0 & \rho_{33} \end{pmatrix} \quad \text{(S27)}$$

which already indicates that the only source of entanglement in the system are the coherences between states $|1\rangle$ and $|2\rangle$. Note that if $\rho_{12} = 0$, the density matrix is diagonal, and the state of the QEs becomes an classical probabilistic mixture without and the emitters are completely unentangled.

The literature characterizing the quantum state of bipartite systems is vast [5–8]. Here, inspired in Ref. [9], we characterize the quantum state described by Eq. (S27) by considering different physical magnitudes. On the one hand, the classical/quantum character of the state can be measured through its Purity and/or Linear Entropy

$$P(\rho) = \text{tr}\{\rho^2\} = \rho_{00}^2 + \rho_{11}^2 + \rho_{22}^2 + \rho_{33}^2 + 2|\rho_{12}|^2, \quad \text{(S28)}$$
$$S_L(\rho) = \frac{4}{3}[1 - \text{tr}\{\rho^2\}] = \frac{4}{3}[1 - (\rho_{00}^2 + \rho_{11}^2 + \rho_{22}^2 + \rho_{33}^2) - 2|\rho_{12}|^2]. \quad \text{(S29)}$$

For a pure quantum state, $P = 1$ and $S_L = 0$, while the maximum Entropy and minimum Purity are $S_L = 1$ and $P = 1/4$, respectively. On the other hand, we will consider two different entanglement witnesses, the Negativity and the Wootters Concurrencence, whose expressions for our system are

$$N(\rho) = \max\{0, \sqrt{(\rho_{33} - \rho_{00})^2 + 4|\rho_{12}|^2} - (\rho_{00} + \rho_{33})\}, \quad \text{(S30)}$$
$$C(\rho) = 2\max\{0, |\rho_{12}| - \sqrt{\rho_{00}\rho_{33}}\}. \quad \text{(S31)}$$

Note that these two magnitudes are maximal, $N = C = 1$ when the population of the ground and double-excited states vanishes and the coherence of the 1-excitation manifold is maximal, $|\rho_{12}| = 1/2$. The resulting density matrix in Eq. (S27) corresponds to the so-called Bell states, of the form

$$|\psi\rangle_{\text{Bell}} = \frac{1}{\sqrt{2}}[|1\rangle + e^{i\theta}|2\rangle] = \frac{1}{\sqrt{2}}[|e\rangle_1|g\rangle_2 + e^{i\theta}|e\rangle_2|g\rangle_1]. \quad \text{(S32)}$$

which are maximally-entangled, pure quantum states.

Fig. S8 shows the four magnitudes above, evaluated from the density matrix in Eqs. (27), obtained by solving Eq. (S26) parameterized with dyadic Green's function calculations of the enhanced configuration. All of them are rendered against the pumping strength of each QE, normalized to its spontaneous decay rate in log-scale. The top panels show that the bipartite quantum states attainable in the system have a moderate quantum character. Except for the limit of vanishing pumping, where the QEs are in their ground state, $|0\rangle = |g\rangle_1|g\rangle_2$. The bottom panels indicate that, despite its mixed character, the QEs density matrix can present significant entanglement in the regime $P_{1,2} \simeq 10^{-1}\gamma_{1,2}$. As expected, due to the incoherent character of the system excitation, the level of entanglement accessible remains the same if only one of the QEs is pumped.

In the literature [9–11], studies establishing the maximum degree of entanglement attainable for a given Linear Entropy of a bipartite state can be found. For the case of bipartite system, the Concurrence of maximally-entangled- mixed states as a function of their Linear Entropy is given by

$$C_{\max}(S_L) = \begin{cases} \frac{1}{2} + \frac{1}{2}\sqrt{1 - \frac{3}{2}S_L}, & 0 \leq S_L \leq \frac{16}{27}, \\ \sqrt{\frac{4}{3} - \frac{3}{2}S_L}, & \frac{16}{27} \leq S_L. \end{cases} \quad (S33)$$

In Fig. S9, the Concurrencence for the states in Fig. S8 is normalized to its maximum value given the quantum/mixed character of the underlying QEs state. We can observe that, in the conditions yielding largest entanglement, the Concurrencence is roughly 60% the maximum attainable. Note that the Purity and Linear Entropy depend severely on the driving scheme, which allows us to conclude that much larger entanglement levels could be obtained with more complex preparation strategies [12].

Calculations for the Concurrencence and negativity of the QEs on top of a flat metal surface and in vacuum yield $N = C = 0$. In these two cases, the QE pair is completely unentangled, which means that the metasurface actually generates it. To compare the three regimes (enhanced, substrate and vacuum), in Fig. S10 we plot the ratio

$$R = \frac{|\rho_{12}|}{\sqrt{\rho_{00}\rho_{33}}}, \quad (S34)$$

which dictates the behaviour of both Negativity and Concurrence in Eqs. (30)-(31). We can observe that this ratio is lower than 1 for all pumping strengths in the center and right panels, which correspond to the flat metal surface and vacuum, respectively. This explains why both $N$ and $C$ vanish in these two cases. On the contrary, in the left panel, $R$ reaches values as large as 10, which yields finite values of both Negativity and Concurrence. These panels demonstrate that, in striking difference, the two QEs become entangled when integrated with enhanced configuration, while their quantum state is completely unentangled and separable in the other platforms.

**Note 5 Steady-State Density-Matrix for Three QEs**

For the case of 3 identical QEs ($\omega_1 = \omega_2 = \omega_3 = \omega_0$), and labelling the basis of the Hilbert space as

$$\begin{aligned}
|0\rangle &= |g\rangle_1|g\rangle_2|g\rangle_3 = |ggg\rangle, & |4\rangle &= |e\rangle_1|e\rangle_2|g\rangle_3 = |eeg\rangle, \\
|1\rangle &= |e\rangle_1|g\rangle_2|g\rangle_3 = |egg\rangle, & |5\rangle &= |e\rangle_1|g\rangle_2|e\rangle_3 = |ege\rangle, \\
|2\rangle &= |g\rangle_1|e\rangle_2|g\rangle_3 = |geg\rangle, & |6\rangle &= |g\rangle_1|e\rangle_2|e\rangle_3 = |gee\rangle, \\
|3\rangle &= |g\rangle_1|g\rangle_2|e\rangle_3 = |gge\rangle, & |7\rangle &= |e\rangle_1|e\rangle_2|e\rangle_3 = |eee\rangle,
\end{aligned} \quad (S35)$$

Eqs. (24) and (25) yield a system of 20 linear equations (only 19 of them are independent). The steady-state equations for the populations have the form

$$\dot{\rho}_{00} = 0 = -(P_1 + P_2 + P_3)\rho_{00} + \gamma_{11}\rho_{11} + \gamma_{22}\rho_{22} + \gamma_{33}\rho_{33} + \gamma_{12}(\rho_{12} + \rho_{12}^*) + \gamma_{13}(\rho_{13} + \rho_{13}^*) \\ + \gamma_{23}(\rho_{23} + \rho_{23}^*),$$

$$\dot{\rho}_{11} = 0 = P_1\rho_{00} - (\gamma_{11} + P_2 + P_3)\rho_{11} + \gamma_{22}\rho_{44} + \gamma_{33}\rho_{55} + \left(ig_{12} - \frac{\gamma_{12}}{2}\right)\rho_{12} - \left(ig_{12} + \frac{\gamma_{12}}{2}\right)\rho_{12}^* \\ + \left(ig_{13} - \frac{\gamma_{13}}{2}\right)\rho_{13} - \left(ig_{13} + \frac{\gamma_{13}}{2}\right)\rho_{13}^* + \gamma_{23}\rho_{45} + \gamma_{23}\rho_{45}^*,$$

$$\dot{\rho}_{22} = 0 = P_2\rho_{00} - (P_1 + \gamma_{22} + P_3)\rho_{22} + \gamma_{11}\rho_{44} + \gamma_{33}\rho_{66} - \left(ig_{12} + \frac{\gamma_{12}}{2}\right)\rho_{12} + \left(ig_{12} - \frac{\gamma_{12}}{2}\right)\rho_{12}^* \\ + \left(ig_{23} - \frac{\gamma_{23}}{2}\right)\rho_{23} - \left(ig_{23} + \frac{\gamma_{23}}{2}\right)\rho_{23}^* + \gamma_{13}\rho_{46} + \gamma_{13}\rho_{46}^*,$$

$$\dot{\rho}_{33} = 0 = P_3\rho_{00} - (P_1 + P_2 + \gamma_{33})\rho_{33} + \gamma_{11}\rho_{55} + \gamma_{22}\rho_{66} - \left(ig_{13} + \frac{\gamma_{13}}{2}\right)\rho_{13} + \left(ig_{13} - \frac{\gamma_{13}}{2}\right)\rho_{13}^* \\ - \left(ig_{23} + \frac{\gamma_{23}}{2}\right)\rho_{23} + \left(ig_{23} - \frac{\gamma_{23}}{2}\right)\rho_{23}^* + \gamma_{12}\rho_{56} + \gamma_{12}\rho_{56}^*,$$

$$\dot{\rho}_{44} = 0 = P_2\rho_{11} + P_1\rho_{22} - (\gamma_{11} + \gamma_{22} + P_3)\rho_{44} + \gamma_{33}\rho_{77} + \left(ig_{23} - \frac{\gamma_{23}}{2}\right)\rho_{45} - \left(ig_{23} + \frac{\gamma_{23}}{2}\right)\rho_{45}^* \\ + \left(ig_{13} - \frac{\gamma_{23}}{2}\right)\rho_{46} - \left(ig_{13} + \frac{\gamma_{23}}{2}\right)\rho_{46}^*,$$

$$\dot{\rho}_{55} = 0 = P_3\rho_{11} + P_1\rho_{33} - (\gamma_{11} + P_2 + \gamma_{33})\rho_{55} + \gamma_{22}\rho_{77} - \left(ig_{23} + \frac{\gamma_{23}}{2}\right)\rho_{45} + \left(ig_{23} - \frac{\gamma_{23}}{2}\right)\rho_{45}^* \\ + \left(ig_{12} - \frac{\gamma_{12}}{2}\right)\rho_{56} - \left(ig_{12} + \frac{\gamma_{12}}{2}\right)\rho_{56}^*,$$

$$\dot{\rho}_{66} = 0 = P_3\rho_{22} + P_2\rho_{33} - (P_1 + \gamma_{22} + \gamma_{33})\rho_{66} + \gamma_{11}\rho_{77} - \left(ig_{13} + \frac{\gamma_{13}}{2}\right)\rho_{46} + \left(ig_{13} - \frac{\gamma_{13}}{2}\right)\rho_{46}^* \\ - \left(ig_{12} + \frac{\gamma_{12}}{2}\right)\rho_{56} + \left(ig_{12} - \frac{\gamma_{12}}{2}\right)\rho_{56}^*,$$

$$\dot{\rho}_{77} = 0 = P_3\rho_{44} + P_2\rho_{55} + P_1\rho_{66} - (\gamma_{11} + \gamma_{22} + \gamma_{33})\rho_{77}, \quad (S36)$$

while the equations for the coherences (non-diagonal terms of the density matrix) read

$$\dot{\rho}_{12} = 0 = \left(ig_{12} - \frac{\gamma_{12}}{2}\right)\rho_{11} - \left(ig_{12} + \frac{\gamma_{12}}{2}\right)\rho_{22} + \gamma_{12}\rho_{44} - \left(\frac{P_1 + \gamma_{11} + P_2 + \gamma_{22}}{2} + P_3\right)\rho_{12} \\ + \left(ig_{23} - \frac{\gamma_{23}}{2}\right)\rho_{13} - \left(ig_{13} + \frac{\gamma_{13}}{2}\right)\rho_{23}^* + \gamma_{13}\rho_{45}^* + \gamma_{23}\rho_{46} + \gamma_{33}\rho_{56},$$

$$\dot{\rho}_{12}^* = 0 = -\left(ig_{12} + \frac{\gamma_{12}}{2}\right)\rho_{11} + \left(ig_{12} - \frac{\gamma_{12}}{2}\right)\rho_{22} + \gamma_{12}\rho_{44} - \left(\frac{P_1 + \gamma_{11} + P_2 + \gamma_{22}}{2} + P_3\right)\rho_{12}^*$$
$$- \left(ig_{23} + \frac{\gamma_{23}}{2}\right)\rho_{13}^* + \left(ig_{13} - \frac{\gamma_{13}}{2}\right)\rho_{23} + \gamma_{13}\rho_{45} + \gamma_{23}\rho_{46}^* + \gamma_{33}\rho_{56}^*,$$

$$\dot{\rho}_{13} = 0 = \left(ig_{13} - \frac{\gamma_{13}}{2}\right)\rho_{11} - \left(ig_{13} + \frac{\gamma_{13}}{2}\right)\rho_{33} + \gamma_{13}\rho_{55} + \left(ig_{23} - \frac{\gamma_{23}}{2}\right)\rho_{12}$$
$$- \left(\frac{P_1 + \gamma_{11} + P_3 + \gamma_{33}}{2} + P_2\right)\rho_{13} - \left(ig_{12} + \frac{\gamma_{12}}{2}\right)\rho_{23} + \gamma_{12}\rho_{45} + \gamma_{22}\rho_{46} + \gamma_{23}\rho_{56},$$

$$\dot{\rho}_{13}^* = 0 = -\left(ig_{13} + \frac{\gamma_{13}}{2}\right)\rho_{11} + \left(ig_{13} - \frac{\gamma_{13}}{2}\right)\rho_{33} + \gamma_{13}\rho_{55} - \left(ig_{23} + \frac{\gamma_{23}}{2}\right)\rho_{12}^*$$
$$- \left(\frac{P_1 + \gamma_{11} + P_3 + \gamma_{33}}{2} + P_2\right)\rho_{13}^* + \left(ig_{12} - \frac{\gamma_{12}}{2}\right)\rho_{23}^* + \gamma_{12}\rho_{45}^* + \gamma_{22}\rho_{46}^* + \gamma_{23}\rho_{56}^*,$$

$$\dot{\rho}_{23} = 0 = \left(ig_{23} - \frac{\gamma_{23}}{2}\right)\rho_{22} - \left(ig_{23} + \frac{\gamma_{23}}{2}\right)\rho_{33} + \gamma_{23}\rho_{66} + \left(ig_{13} - \frac{\gamma_{13}}{2}\right)\rho_{12}^* - \left(ig_{12} + \frac{\gamma_{12}}{2}\right)\rho_{13}$$
$$- \left(P_1 + \frac{P_2 + \gamma_{22} + P_3 + \gamma_{33}}{2}\right)\rho_{23} + \gamma_{11}\rho_{45} + \gamma_{12}\rho_{46} + \gamma_{13}\rho_{56}^*,$$

$$\dot{\rho}_{23}^* = 0 = -\left(ig_{23} + \frac{\gamma_{23}}{2}\right)\rho_{22} + \left(ig_{23} - \frac{\gamma_{23}}{2}\right)\rho_{33} + \gamma_{23}\rho_{66} - \left(ig_{13} + \frac{\gamma_{13}}{2}\right)\rho_{12} + \left(ig_{12} - \frac{\gamma_{12}}{2}\right)\rho_{13}^*$$
$$- \left(P_1 + \frac{P_2 + \gamma_{22} + P_3 + \gamma_{33}}{2}\right)\rho_{23}^* + \gamma_{11}\rho_{45}^* + \gamma_{12}\rho_{46}^* + \gamma_{13}\rho_{56},$$

$$\dot{\rho}_{45} = 0 = \left(ig_{23} - \frac{\gamma_{23}}{2}\right)\rho_{44} - \left(ig_{23} + \frac{\gamma_{23}}{2}\right)\rho_{55} + \gamma_{23}\rho_{77} + P_1\rho_{23} - \left(\gamma_{11} + \frac{P_2 + \gamma_{22} + P_3 + \gamma_{33}}{2}\right)\rho_{45}$$
$$+ \left(ig_{12} - \frac{\gamma_{12}}{2}\right)\rho_{46} - \left(ig_{13} + \frac{\gamma_{13}}{2}\right)\rho_{56}^*,$$

$$\dot{\rho}_{45}^* = 0 = -\left(ig_{23} + \frac{\gamma_{23}}{2}\right)\rho_{44} + \left(ig_{23} - \frac{\gamma_{23}}{2}\right)\rho_{55} + \gamma_{23}\rho_{77} + P_1\rho_{13}^*$$
$$- \left(\gamma_{11} + \frac{P_2 + \gamma_{22} + P_3 + \gamma_{33}}{2}\right)\rho_{45}^* - \left(ig_{12} + \frac{\gamma_{12}}{2}\right)\rho_{46}^* + \left(ig_{13} - \frac{\gamma_{13}}{2}\right)\rho_{56},$$

$$\dot{\rho}_{46} = 0 = \left(ig_{13} - \frac{\gamma_{13}}{2}\right)\rho_{44} - \left(ig_{13} + \frac{\gamma_{13}}{2}\right)\rho_{66} + \gamma_{13}\rho_{77} + P_2\rho_{13} + \left(ig_{12} - \frac{\gamma_{12}}{2}\right)\rho_{45}$$
$$- \left(\gamma_{22} + \frac{P_1 + \gamma_{11} + P_3 + \gamma_{33}}{2}\right)\rho_{46} - \left(ig_{13} + \frac{\gamma_{13}}{2}\right)\rho_{56},$$

$$\dot{\rho}_{46}^* = 0 = -\left(ig_{13} + \frac{\gamma_{13}}{2}\right)\rho_{44} + \left(ig_{13} - \frac{\gamma_{13}}{2}\right)\rho_{66} + \gamma_{13}\rho_{77} + P_2\rho_{13}^* - \left(ig_{12} + \frac{\gamma_{12}}{2}\right)\rho_{45}^*$$
$$- \left(\gamma_{22} + \frac{P_1 + \gamma_{11} + P_3 + \gamma_{33}}{2}\right)\rho_{46}^* + \left(ig_{13} - \frac{\gamma_{13}}{2}\right)\rho_{56}^*,$$

$$\dot{\rho}_{56} = 0 = \left(ig_{12} - \frac{\gamma_{12}}{2}\right)\rho_{55} - \left(ig_{12} + \frac{\gamma_{12}}{2}\right)\rho_{66} + \gamma_{12}\rho_{77} + P_3\rho_{12} + \left(ig_{13} - \frac{\gamma_{13}}{2}\right)\rho_{45}^*$$
$$- \left(ig_{23} + \frac{\gamma_{23}}{2}\right)\rho_{46} - \left(\gamma_{33} + \frac{P_1 + \gamma_{11} + P_2 + \gamma_{22}}{2}\right)\rho_{56},$$

$$\dot{\rho}_{56}^* = 0 = -\left(ig_{12} + \frac{\gamma_{12}}{2}\right)\rho_{55} + \left(ig_{12} - \frac{\gamma_{12}}{2}\right)\rho_{66} + \gamma_{12}\rho_{77} + P_3\rho_{12}^* - \left(ig_{13} + \frac{\gamma_{13}}{2}\right)\rho_{45}$$
$$+ \left(ig_{23} - \frac{\gamma_{23}}{2}\right)\rho_{46}^* - \left(\gamma_{33} + \frac{P_1 + \gamma_{11} + P_2 + \gamma_{22}}{2}\right)\rho_{56}^*,$$
(S37)

and the final equation, given by the trace, reads
$$tr\{\rho\} = 1 = \rho_{00} + \rho_{11} + \rho_{22} + \rho_{33} + \rho_{44} + \rho_{55} + \rho_{66} + \rho_{77},$$
(S38)

The resulting density matrix has the form

$$\rho = \begin{pmatrix} \rho_{00} & 0 & 0 & 0 & 0 & 0 & 0 & 0 \\ 0 & \rho_{11} & \rho_{12} & \rho_{13} & 0 & 0 & 0 & 0 \\ 0 & \rho_{12}^* & \rho_{22} & \rho_{23} & 0 & 0 & 0 & 0 \\ 0 & \rho_{13}^* & \rho_{23}^* & \rho_{33} & 0 & 0 & 0 & 0 \\ 0 & 0 & 0 & 0 & \rho_{44} & \rho_{45} & \rho_{46} & 0 \\ 0 & 0 & 0 & 0 & \rho_{45}^* & \rho_{55} & \rho_{56} & 0 \\ 0 & 0 & 0 & 0 & \rho_{46}^* & \rho_{56}^* & \rho_{66} & 0 \\ 0 & 0 & 0 & 0 & 0 & 0 & 0 & \rho_{77} \end{pmatrix}$$
(S39)

We also construct the density matrices resulting from the tracing out of one of the qubits, which, using the site basis can be expressed as

$$\rho^{23} = tr_1\{\rho\} \quad \text{with } \rho^{23}_{jk,j'k'} = \Sigma_{i=e,g} |jk\rangle\langle ijk|\rho|i'j'k'\rangle\langle j'k'|,$$
$$\rho^{13} = tr_2\{\rho\} \quad \text{with } \rho^{13}_{ik,i'k'} = \Sigma_{j=e,g} |ik\rangle\langle ijk|\rho|i'j'k'\rangle\langle i'k'|,$$
$$\rho^{12} = tr_3\{\rho\} \quad \text{with } \rho^{12}_{ij,i'j'} = \Sigma_{k=e,g} |ij\rangle\langle ijk|\rho|i'j'k\rangle\langle i'j'|.$$
(S40)

For our system, we get

$$\rho^{23} = \begin{pmatrix} \rho_{00} + \rho_{11} & 0 & 0 & 0 \\ 0 & \rho_{22} + \rho_{44} & \rho_{23} + \rho_{45} & 0 \\ 0 & \rho_{23}^* + \rho_{45}^* & \rho_{33} + \rho_{55} & 0 \\ 0 & 0 & 0 & \rho_{66} + \rho_{77} \end{pmatrix}$$

$$\rho^{13} = \begin{pmatrix} \rho_{00} + \rho_{22} & 0 & 0 & 0 \\ 0 & \rho_{11} + \rho_{44} & \rho_{13} + \rho_{46} & 0 \\ 0 & \rho_{13}^* + \rho_{46}^* & \rho_{33} + \rho_{66} & 0 \\ 0 & 0 & 0 & \rho_{55} + \rho_{77} \end{pmatrix}$$

$$\rho^{23} = \begin{pmatrix} \rho_{00} + \rho_{33} & 0 & 0 & 0 \\ 0 & \rho_{11} + \rho_{55} & \rho_{12} + \rho_{56} & 0 \\ 0 & \rho_{12}^* + \rho_{56}^* & \rho_{22} + \rho_{66} & 0 \\ 0 & 0 & 0 & \rho_{44} + \rho_{77} \end{pmatrix}$$

$$\rho^1 = \begin{pmatrix} \rho_{00} + \rho_{11} + \rho_{22} + \rho_{44} & 0 \\ 0 & \rho_{33} + \rho_{55} + \rho_{66} + \rho_{77} \end{pmatrix}$$
(S41)

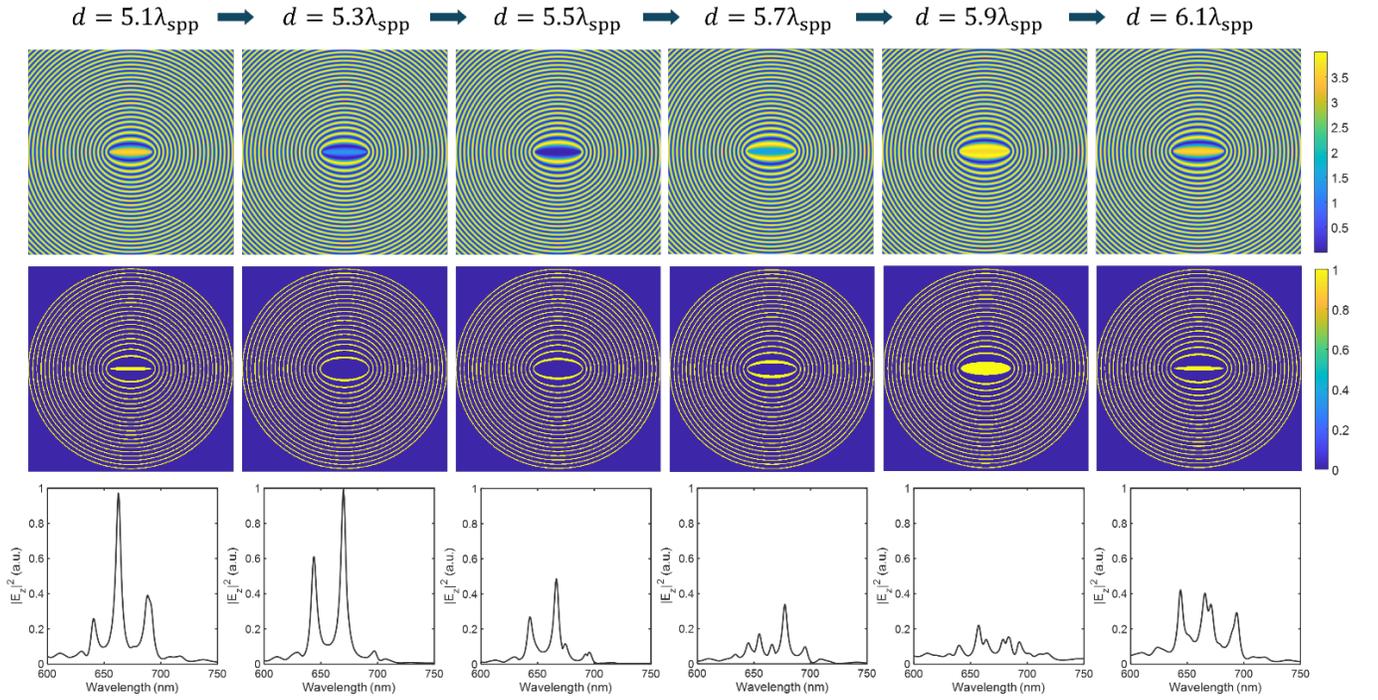

**Fig. S1 The evolution of QE–QE coupling for different QE separation distances.** The coupling behaviour is investigated for different distances between the two QEs, ranging from $d = 5.1\lambda_{\text{spp}}$ to $d = 6.1\lambda_{\text{spp}}$. Top row shows the intensity interference patterns generated by using Eq. (S4) for the operating wavelength of 670 nm and the filling ration of 0.25. Middle row shows the corresponding (binarized) design patterns obtained using the same threshold. Bottom row represents the field intensity spectra at the acceptor point normalized to the intensity maximum obtained with $d = 5.3\lambda_{spp}$. As the separation is tuned, the amplitude and position of resonance peaks shift.

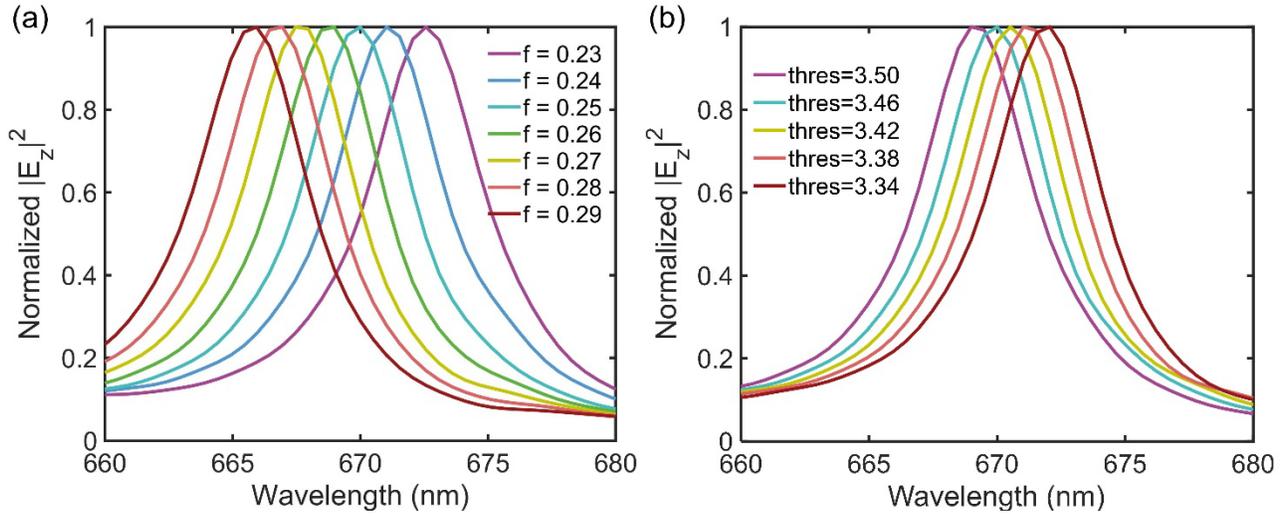

**Fig. S2 Precise control of the resonance wavelength by varying the metasurface filling ratio and threshold parameter.** (a) Resonance wavelength tuning by varying the metasurface filling ratio $f$ (from 0.25 to 0.29). In this case, the preset filling ratio $\xi$ is set equal to the real structure filling ratio $f$, such that changing $f$ also change $d$. (b) Resonance tuning by varying the threshold parameter used to define the width of holographic ridges (therefore also changing the structure filling ratio $f$), while keeping preset filling ratio $\xi = 0.25$. Spectra are calculated for a fixed QE separation $d = 5.3\lambda_{\text{spp}}$. In both cases, the resonance wavelength can be continuously tuned, demonstrating the flexibility of the holographic metasurface design.

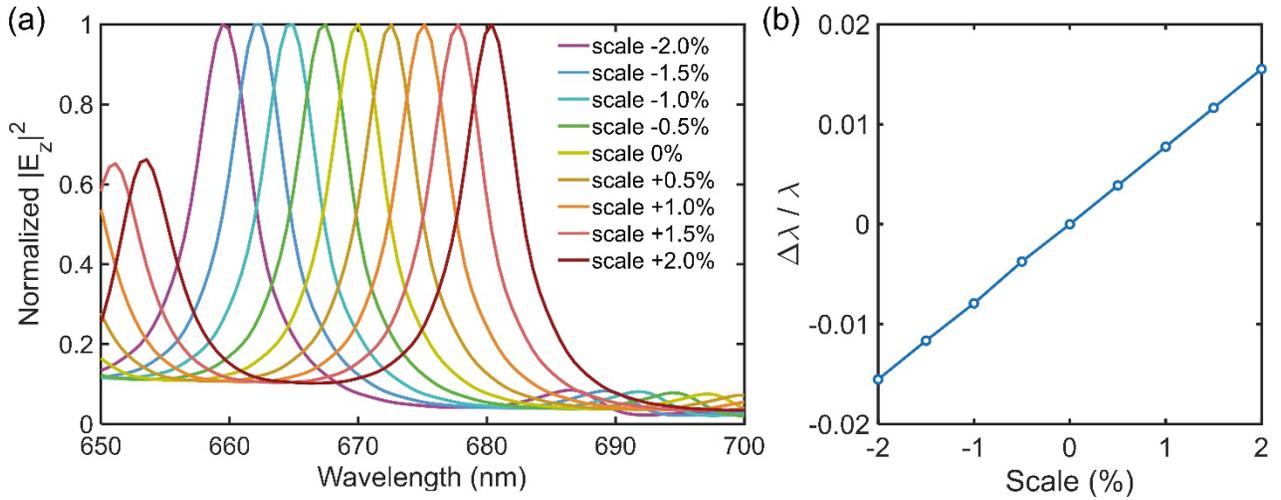

**Fig. S3 Resonance tuning by geometric in-plane scaling of the holographic metasurface.** The on-chip metasurface is uniformly scaled in the lateral plane by −2.0% to +2.0%, with the thickness kept unchanged. (a) The resonance wavelength shifts systematically to shorter (longer) wavelengths under contraction (expansion). (b) Relative resonance shift $\Delta\lambda/\lambda$ as a function of the geometric scaling factor. The approximately linear dependence confirms that the resonance wavelength scales proportionally with the metasurface geometry. This scaling-based approach enables precise tuning of the resonance to a target wavelength without reoptimizing the metasurface by modifying the threshold and/or filling ratio, providing another robust and practical strategy for resonance tuning, thus complementing the approaches introduced in Fig. S2.

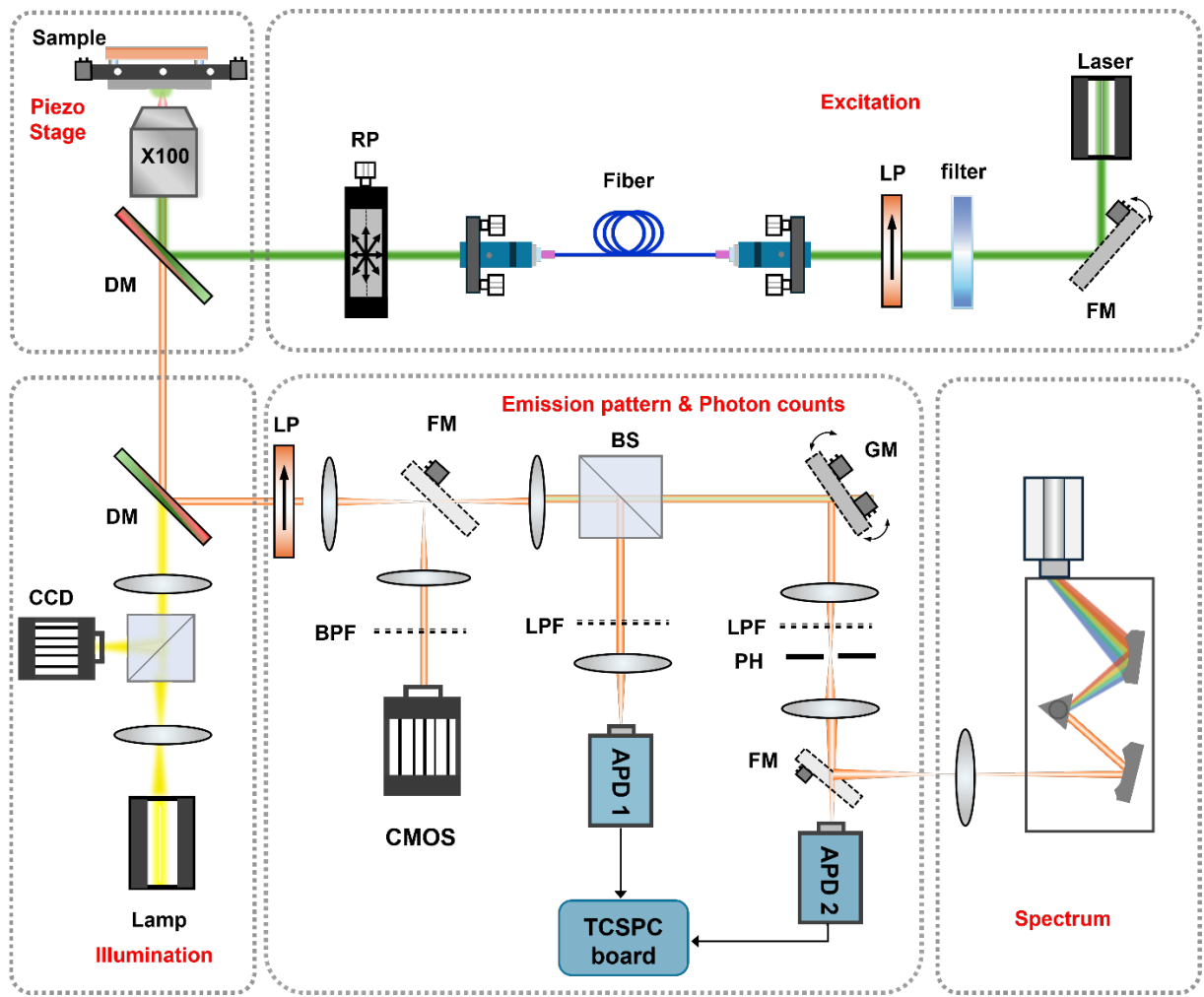

**Fig. S4** Experimental setup for characterizing the samples. A piezo-stage allows for locating NVs-NDs in structures. A 532 nm incident laser is used for exciting NV-NDs. Illumination module enables finding the fabricated microscale structures. Characterization module is used for measuring emission patterns, photon counts, and spectrum. CW: continuous wave, RP: radial polarization converter, PBS: polarized beam splitter, PH: pinhole, DM: dichroic mirror, LP: linear polarizer, LPF: 550 nm long pass filter, BPF: 670 nm ± 5 nm. FM: flip mirror, GM: galvanometric mirror. APD: avalanche photodiode.

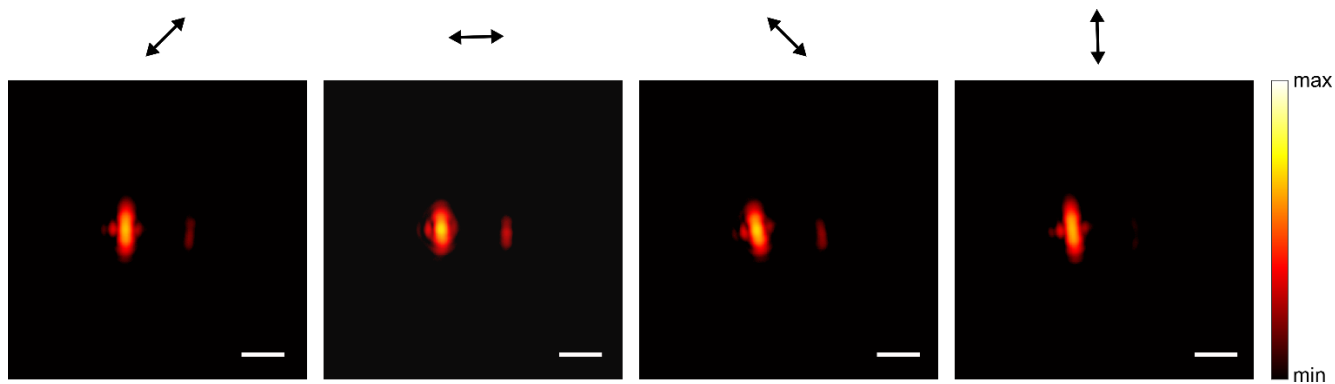

**Fig. S5** Fluorescence images of the elliptic enhanced configuration with a rotation of a linear polarizer, showing that the dominant polarization of out-of-plane scattered fields at the acceptor is the $x$ polarization, which is aligned with the line connecting the two QEs and the oscillation direction inside the on-chip cavity.

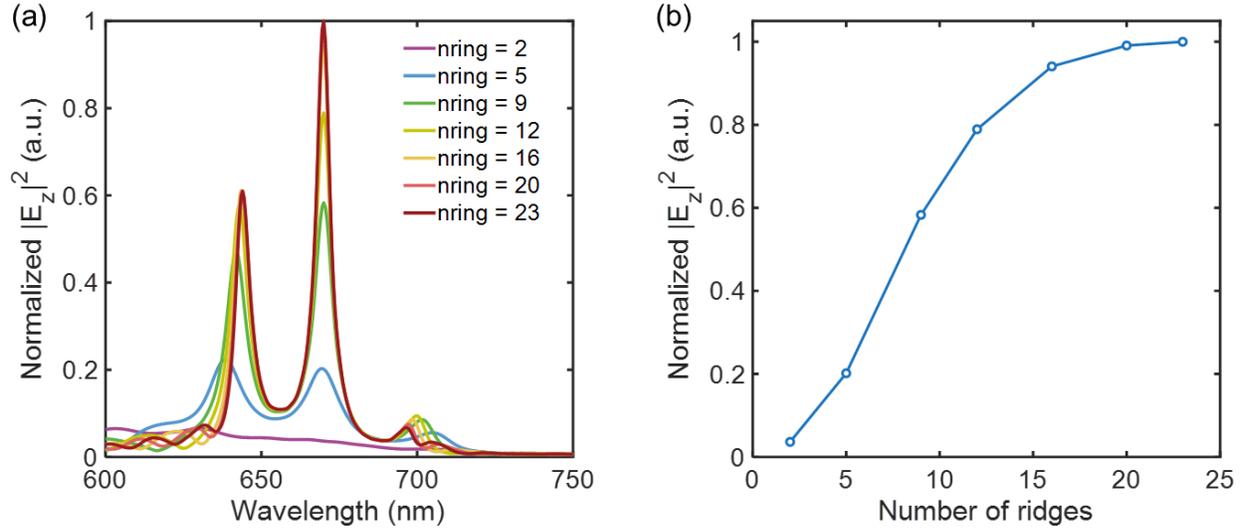

**Fig. S6 Dependence of the metasurface size on the spectral field enhancement.** (a) Spectral field enhancement of holographic metasurfaces with different numbers of nanostructure ridges ($n_{\text{ring}}$). (b) Normalized $|E_z|^2$ evaluated at $\lambda = 670$ nm as a function of the number of ridges. The field enhancement increases with the metasurface sizes and gradually saturates for $n_{\text{ring}} \gtrsim 15$–20. With increasing metasurface size, the resonance peaks become more pronounced, indicating an increasing effective resonator length. The effective resonator length is estimated from the extracted free spectral range (FSR) using $\Delta\lambda_{\text{FSR}} = \frac{\lambda^2}{n_g L}$, where $\lambda$ is the resonance wavelength, $n_g$ is the SPP group index, and $L$ is the round-trip length of the equivalent cavity. With $\Delta\lambda_{\text{FSR}} = 26$ nm, $\lambda = 670$ nm and $n_g = 1.62$, we obtain $L \approx 10.6$, i.e., approximately 13 ridges. The metasurface used in simulations and experiments ($n_{\text{ring}} = 20$) ensures robust field enhancement while remaining insensitive to further increases in size.

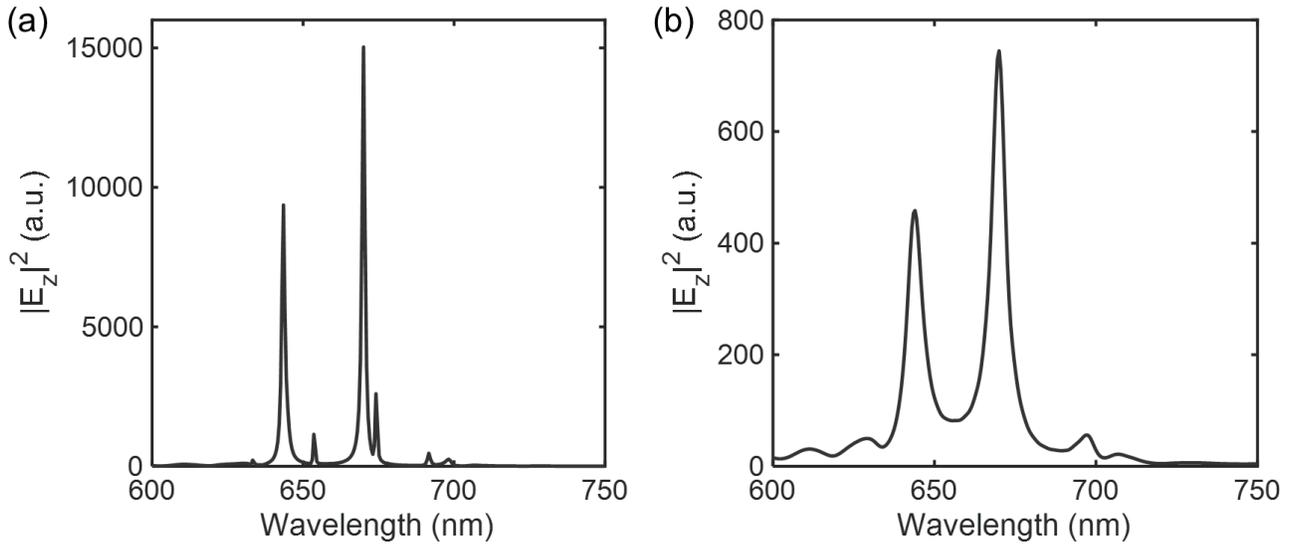

**Fig. S7 Effect of material losses on the metasurface-mediated field enhancement.** Spectra calculated for identical configurations without (a) and with (b) material losses. In low-loss case (a), the imaginary part of the silver permittivity is strongly reduced (Im $\varepsilon \approx 0.1$), effectively suppressing ohmic dissipation. In the real case (b), the full complex permittivity of silver is used. The loss-suppressed structure exhibits significantly higher peak enhancement and sharper spectral features compared to the realistic lossy case. This comparison demonstrates the potential for further enhancement of the interaction by making use of low-loss materials.

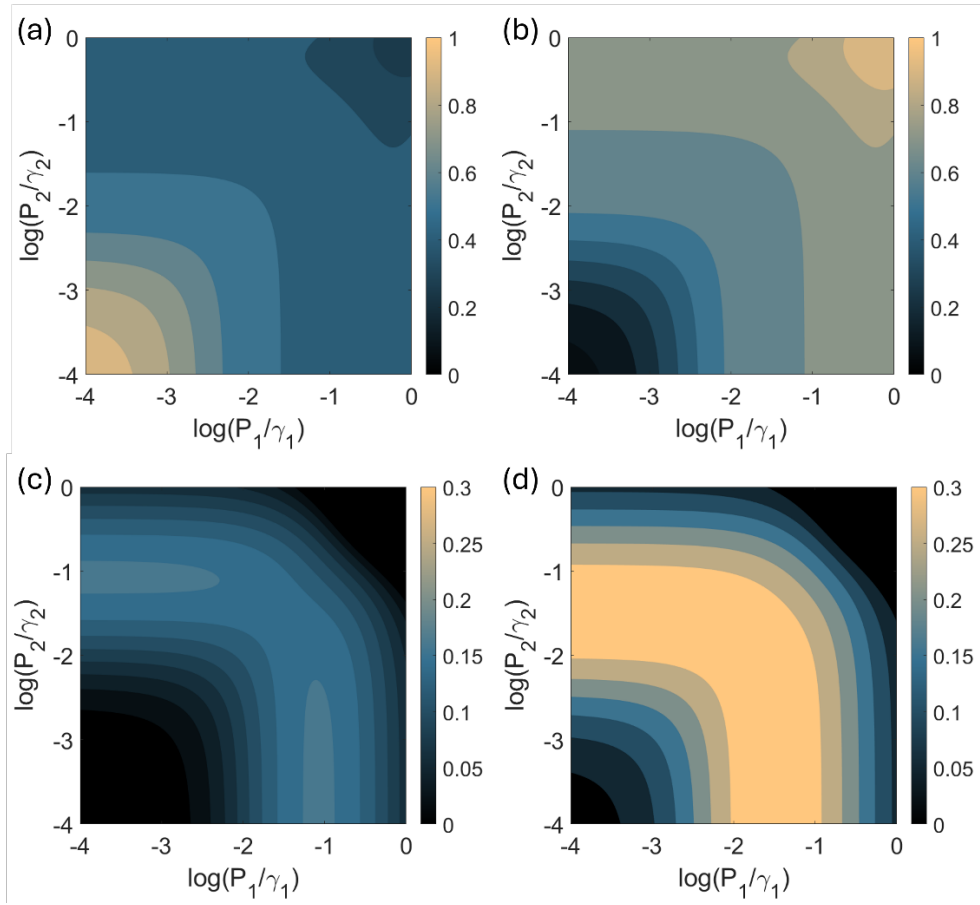

**Fig. S8** (a) Purity, (b) Linear Entropy, (c) Negativity, and (d) Concurrence for the density matrix in Eqs. (27), obtained from the solution of Eqs. (26) parameterized with dyadic Green's function calculations from simulations. All magnitudes are rendered against the pumping strength of each QE, normalized to its spontaneous decay rate.

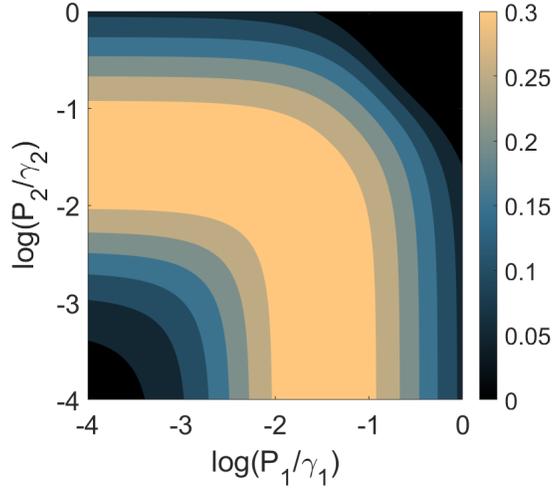

**Fig. S9** Concurrence normalized to its maximum value given the Linear Entropy of the quantum state of the QE pair. The ratio is rendered against the pumping strength of each QE, normalized to its spontaneous decay rate.

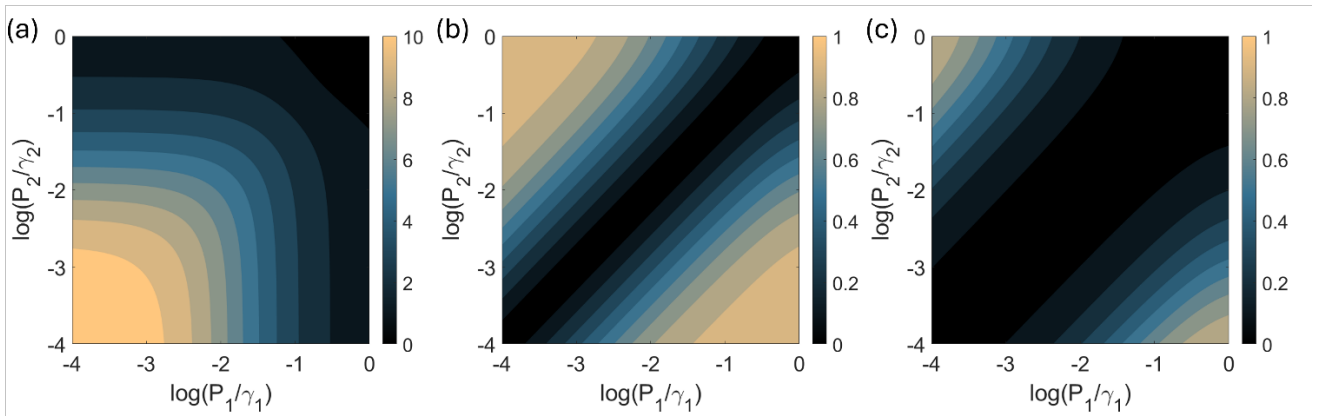

**Fig. S10** Ratio between 1-excitation manifold coherences and the square root of the ground and double-excited populations, $R$, for three different systems: (a) enhanced configuration, (b) flat substrate, and (c) vacuum. In all cases, Eqs. (26) have been parametrized with simulation.

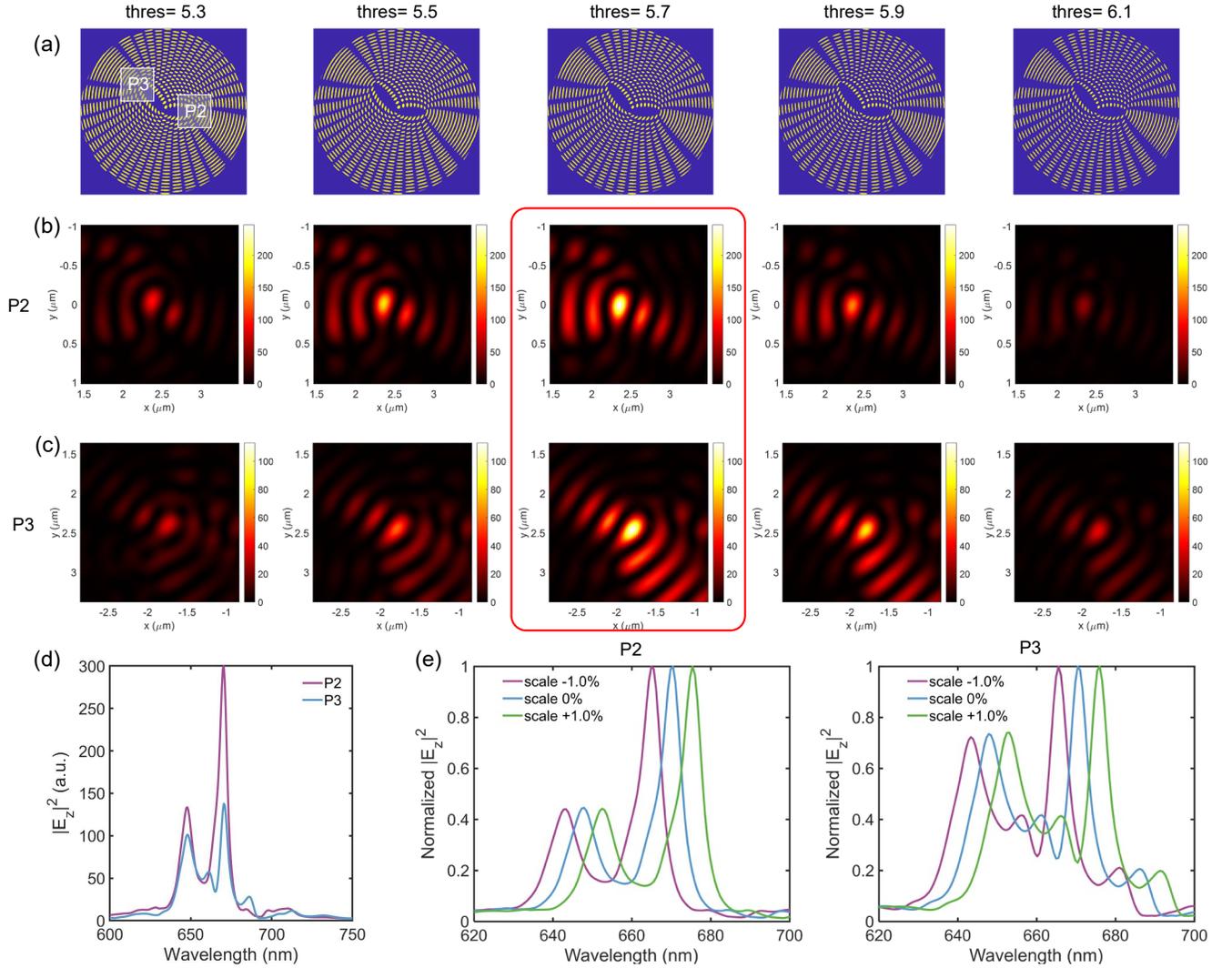

**Fig. S11 Optimization of holographic metasurfaces for 3 QE coupling (Case 1).** (a) Holographic metasurfaces designed for 3 QE coupling with an increasing threshold parameter from left to right. (b, c) Simulated far-field intensity distributions evaluated at detection planes centered at Position 2 (P2) and Position 3 (P3). The optimized overall field enhancement is highlighted by a red box. (d) Spectra extracted at P2 and P3. (e) Resonance manipulation by geometric in-plane scaling of the holographic metasurface indicating similar resonance wavelength tunings for both positions (P2 and P3).

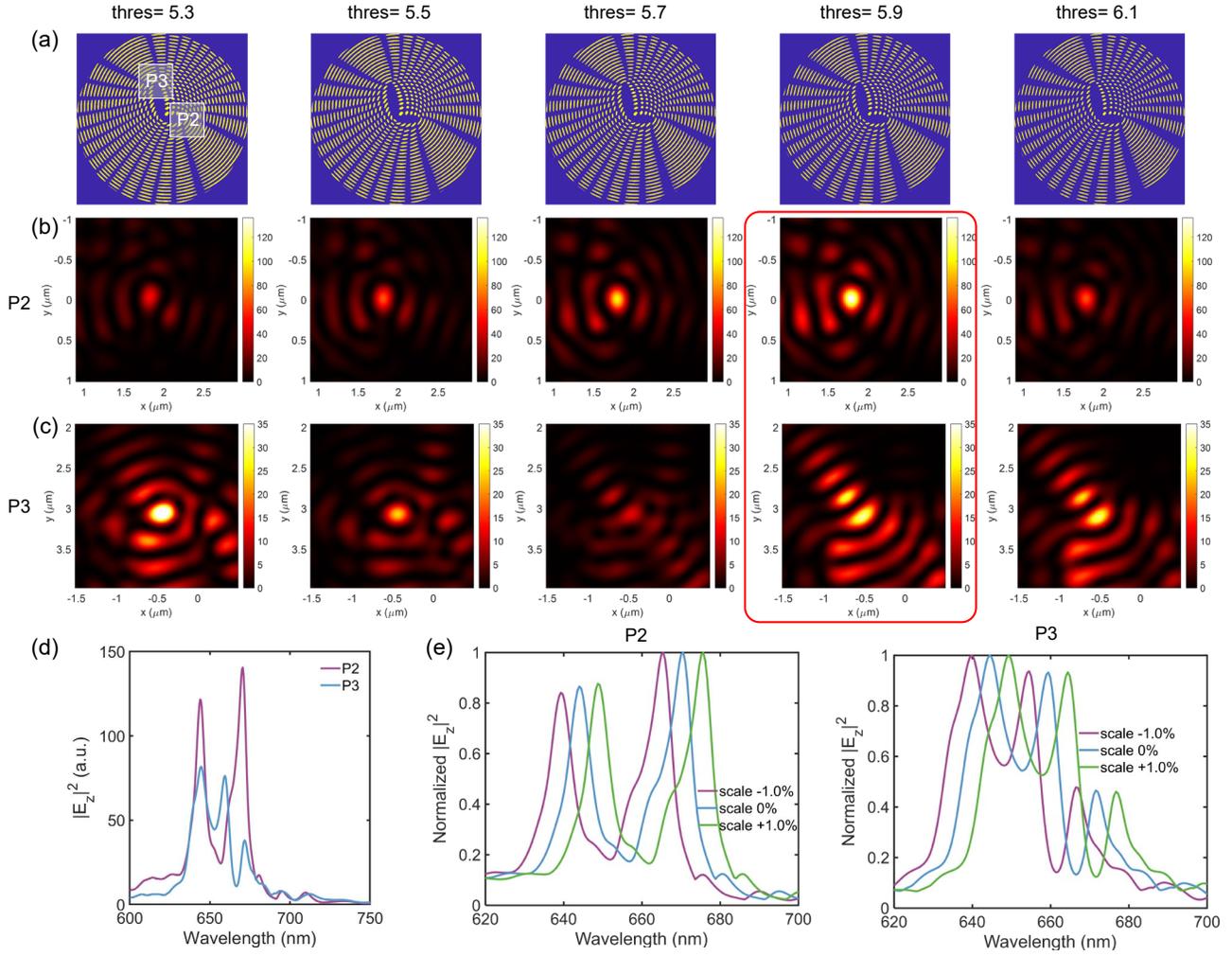

**Fig. S12 Optimization of holographic metasurfaces for 3 QE coupling (Case 2).** (a) Holographic metasurfaces designed for 3 QE coupling with an increasing threshold parameter from left to right. (b, c) Simulated far-field intensity distributions evaluated at detection planes centered at Position 2 (P2) and Position 3 (P3). The optimized overall field enhancement is highlighted by a red box. (d) Spectra extracted at P2 and P3. (e) Resonance manipulation by geometric in-plane scaling of the holographic metasurface indicating similar resonance wavelength tunings for both positions (P2 and P3).